\documentclass[aps,pra,twocolumn,superscriptaddress,nofootinbib]{revtex4-1}
\usepackage{graphicx,graphics,xcolor,braket,amsmath,mathtools,subfigure,bm,bbm,tensor}
\usepackage{hyperref}
\usepackage{cleveref}
\usepackage{stmaryrd}

\crefname{equation}{Eq.}{Eqs.}
\Crefname{equation}{Equation}{Equations}
\crefname{figure}{Fig.}{Figs.}
\Crefname{figure}{Figure}{Figures}
\crefname{section}{Sec.}{Secs.}
\crefname{subsection}{Subsec.}{Subsecs.}
\Crefname{section}{Section}{Sections}
\crefname{appendix}{Appendix}{Appendices}
\Crefname{appendix}{Appendix}{Appendices}

\usepackage{xr-hyper}
\usepackage[normalem]{ulem}

\newcommand{\eqn}[1]{Eq.~(\ref{#1})}

\begin{document}

\title{Doubly nonlinear superconducting qubit}

\author{Dat Thanh Le} \email[]{thanhdat.le@uq.net.au}
\affiliation{ARC Centre for Engineered Quantum System, Department of Physics, University of Queensland, Brisbane, QLD 4072, Australia}
\affiliation{Thang Long Institute of Mathematics and Applied Sciences (TIMAS), Thang Long University, Nghiem Xuan Yem, Hoang Mai, Hanoi 10000, Vietnam}

\author{Arne Grimsmo} 
\affiliation{ARC Centre for Engineered Quantum Systems, School of Physics, The University of Sydney, Sydney, NSW 2006, Australia}

\author{Clemens M\"uller} 
\affiliation{ARC Centre for Engineered Quantum System, Department of Physics, University of Queensland, Brisbane, QLD 4072, Australia}
\affiliation{IBM Research Zurich, 8803 R\"uschlikon, Switzerland}

\author{T. M. Stace} \email[]{stace@physics.uq.edu.au}
\affiliation{ARC Centre for Engineered Quantum System, Department of Physics, University of Queensland, Brisbane, QLD 4072, Australia}

\begin{abstract}

We describe a superconducting circuit consisting of a Josephson junction in parallel with a quantum phase slip wire, which implements a Hamiltonian that is periodic in both charge and flux.  This Hamiltonian is exactly diagonalisable in a double-Bloch band, and the eigenstates are shown to be code states of the Gottesman-Kitaev-Preskill quantum error correcting code.  The eigenspectrum has several critical points, where the linear sensitivity to external charge and flux noise vanishes.  The states at these critical points thus hold promise as qubit states that are insensitive to common external noise sources. 
\end{abstract}

\maketitle

\section{Introduction}

Quantum devices are typically sensitive to noise, which presents the major challenge in developing robust quantum technologies.  
In contrast, digital technologies  rest on the existence of stable  states of matter that  retain classical information over long times.  Fundamentally, this is because stable classical states of matter  embody an error correcting code. For example, ferromagnet domains in hard disks 
 energetically implement a repetition code amongst many coupled electronic spins.  By analogy, it is desirable to engineer quantum systems whose Hamiltonians encode a quantum error correcting code.

One approach to developing robust quantum devices is to design  a `symmetry protected' logical space of nearly degenerate ground states $\{\ket{\bar 0},\ket{\bar 1}\}$ \cite{bonderson2013quasi},  such as the proposed $0$--$\pi$ qubit \cite{kitaev2006protected,Brooks13,Dempster14,Groszkowski2018,Paolo19}, which rejects charge and flux noise. 
Here we {instead} introduce and analyse a simple superconducting circuit with a  set of eigenstates that are robust against noise, without relying on ground state degeneracy.
This device is built from a Josephson junction (JJ) \cite{
Martinis87} and a quantum phase slip (QPS) wire \cite{Mooij05,Mooij06,Astafiev12,deGraaf18}, making its Hamiltonian periodic in both charge and flux. The two junctions in this circuit are dual to each other, by which we name the device the dualmon. We show that the dualmon Hamiltonian is exactly diagonalisable, where there are two quantum numbers each associated to one canonical coordinate, and that typical noise processes commute with the Hamiltonian, affording some symmetry protection to the device.

The energy eigenbasis of the dualmon circuit includes the codewords of the Gottesman-Kitaev-Preskill (GKP) error correcting code \cite{Gottesman01,Travaglione02} which occur at  four critical points in the eigenspectrum: one minimum (ground state), one maximum, and two saddle points.  At these critical points, we find that the device is insensitive (at linear order) to fluctuations in both charge and flux, making the critical points {promising} candidates for robust quantum information storage.  We also show that these results hold when the circuit includes realistic {parasitic} inductance and  capacitance.

The paper is structured as follows. In \cref{sec:ideal}, we analyse the characteristics of the elementary dualmon circuit and its robustness to classical noise. Section \ref{sec:parasitic} takes into account the effect of wire inductance and junction capacitance, which are always present in a realistic circuit. We compute the energy bands of this circuit both analytically and numerically to assess the influence of external flux and charge noise. We also discuss on other possible noise sources in the realistic dualmon circuit. In \cref{sec:waveguide}, we couple  the circuit to a waveguide to implement spectroscopy. The resulting transmission spectrum shows that the interband transition is dependent upon the system state, based on which we propose a means for state initialisation.  The waveguide coupling additionally gives rise to a quantum noise model; however, it is found that the dualmon remains resilient against the induced quantum noise. Afterwards, in \cref{sec:compare} we compare the dualmon with several previously investigated superconducting qubit designs.  We conclude the paper in \cref{sec:conclusion}. Several appendices are attached providing details of calculations described in the main text.

\begin{figure}[t]
    \centering
    \includegraphics{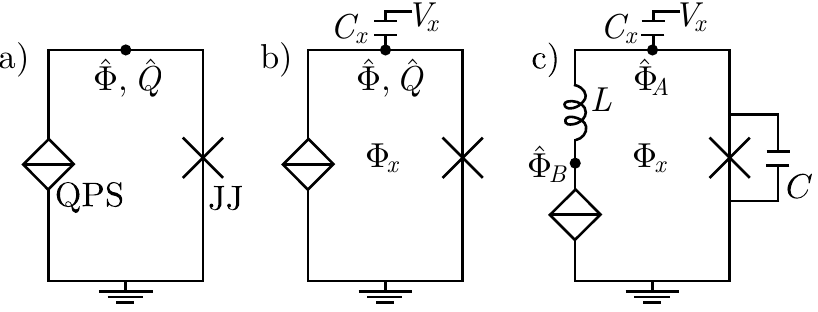}
    \caption{(a) The elementary dualmon circuit of a quantum phase slip (QPS) wire in parallel with a Josephson junction (JJ). (b) Including coupling to external flux $\Phi_x$ and voltage bias $V_x$. 
(c) The realistic dualmon circuit including self inductance $L$, and parasitic capacitance $C$.  }
    \label{Fig:circuit}
\end{figure}

\section{The elementary circuit} \label{sec:ideal}

 \Cref{Fig:circuit}a illustrates the elementary dualmon circuit in which an ideal QPS is in parallel with an ideal JJ, so that there is no parasitic capacitance or inductance. The QPS and JJ are dual circuit elements, with constitutive relations
\mbox{$V_Q = V_c \sin ( 2\pi  {Q}/{(2e)}  )$} and \mbox{$I_J = I_c \sin ( 2\pi {\Phi}/{\Phi_0}  )$} respectively, 
where $V_Q$ is the QPS voltage which depends on the charge,  $Q$, that has flowed through the QPS, $I_J$ is the JJ current which depends on the flux, $\Phi$,  linked by the JJ, and $\Phi_0=h/(2e)$ is the magnetic flux quantum. The QPS and JJ elements are characterised by their critical voltage $V_c$ and critical current $I_c$ respectively.

The quantised circuit is described by the flux  $\hat \Phi$ and the conjugate charge  $\hat Q$. The Hamiltonian for the system is given by \begin{equation}
 {\hat{H}}= -E_Q \cos ( 2\pi \hat{n} ) - E_J \cos (\hat{\phi}), \label{Eq:Ideal_Hamiltonian}
\end{equation} 
where \mbox{$ \hat{n} \!= \! {\hat Q}/{(2e)}$} and \mbox{$\hat \phi \!= \! 2\pi \hat{\Phi}/\Phi_0$} satisfy  \mbox{$[\hat \phi,\hat n]\!=\!i$},  \mbox{$E_Q = {2e V_c}/{(2\pi)}$}, and \mbox{$E_J = { \Phi_0 I_c}/{(2\pi)}$}.
The Hamiltonian is  periodic in both charge and flux, and since  \mbox{$[e^{\pm i\hat\phi},e^{\pm i2\pi \hat n}]=0$}, we find that $[\cos(2\pi \hat n),\cos(\hat \phi)]=0$. The eigenstates of the system $\ket{k,\varphi}$ are therefore characterised by Bloch quantum numbers \mbox{$k\in(-1/2,1/2]$} and \mbox{$\varphi\in(-\pi,\pi]$}, and satisfy dual Bloch relations \mbox{$\braket{  {k,\varphi} | \phi+2\pi}_{\bar\phi}=e^{i 2\pi k }\braket{  {k,\varphi} | \phi}_{\bar\phi}$}  \cite{likharev1985theory}  and 
\mbox{$\braket{  {k,\varphi} | n+1}_{\bar n}=e^{ i\varphi}\braket{  {k,\varphi} | n}_{\bar n}$}, where the subscripts $\bar \phi$ and $\bar n$ distinguish the phase and number bases respectively. The basis $\{ \ket{k,\varphi} \}$ was introduced and analysed in the work by Zak \cite{Zak67}; we will subsequently call it the Zak basis.

\begin{figure}
    \centering
    \includegraphics[height=4.27 cm]{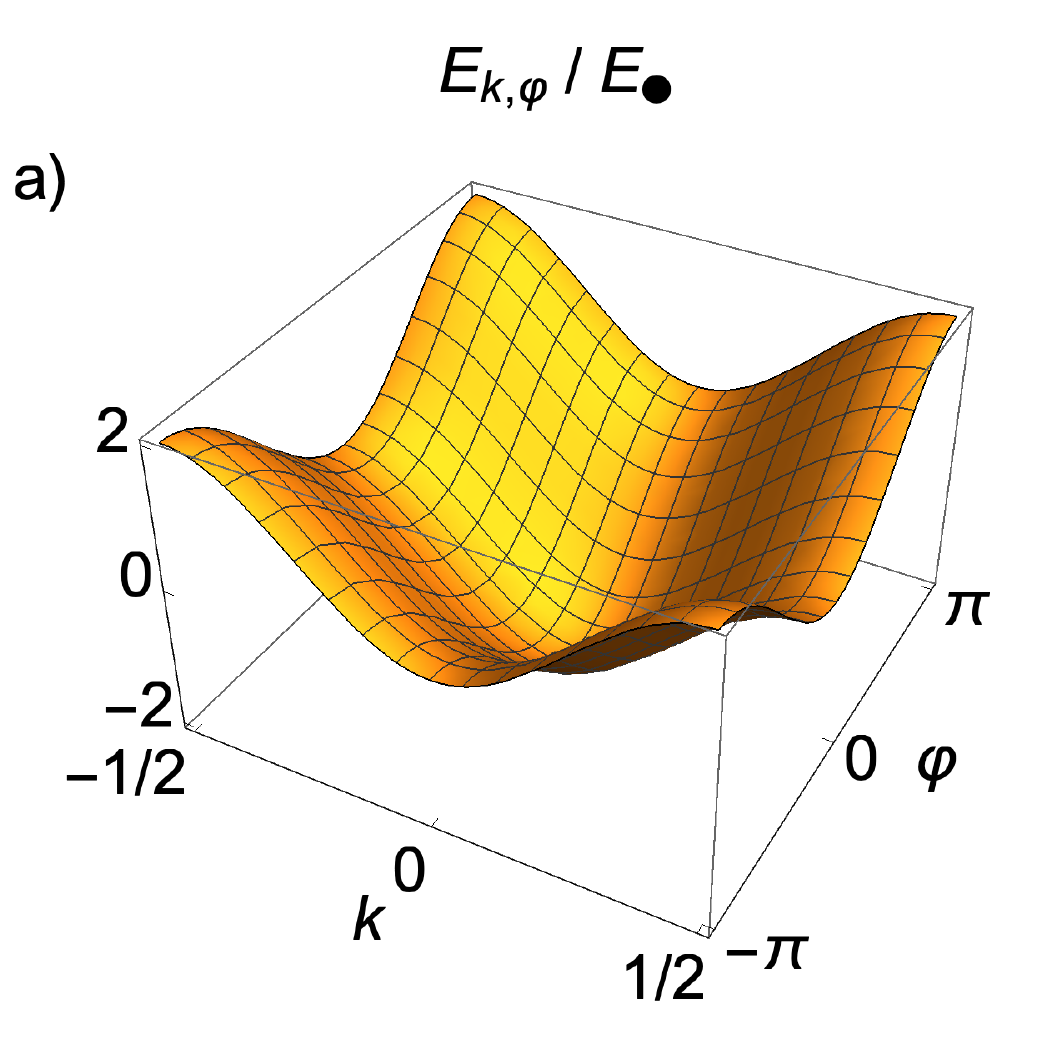}    
    \includegraphics[height=4.27 cm]{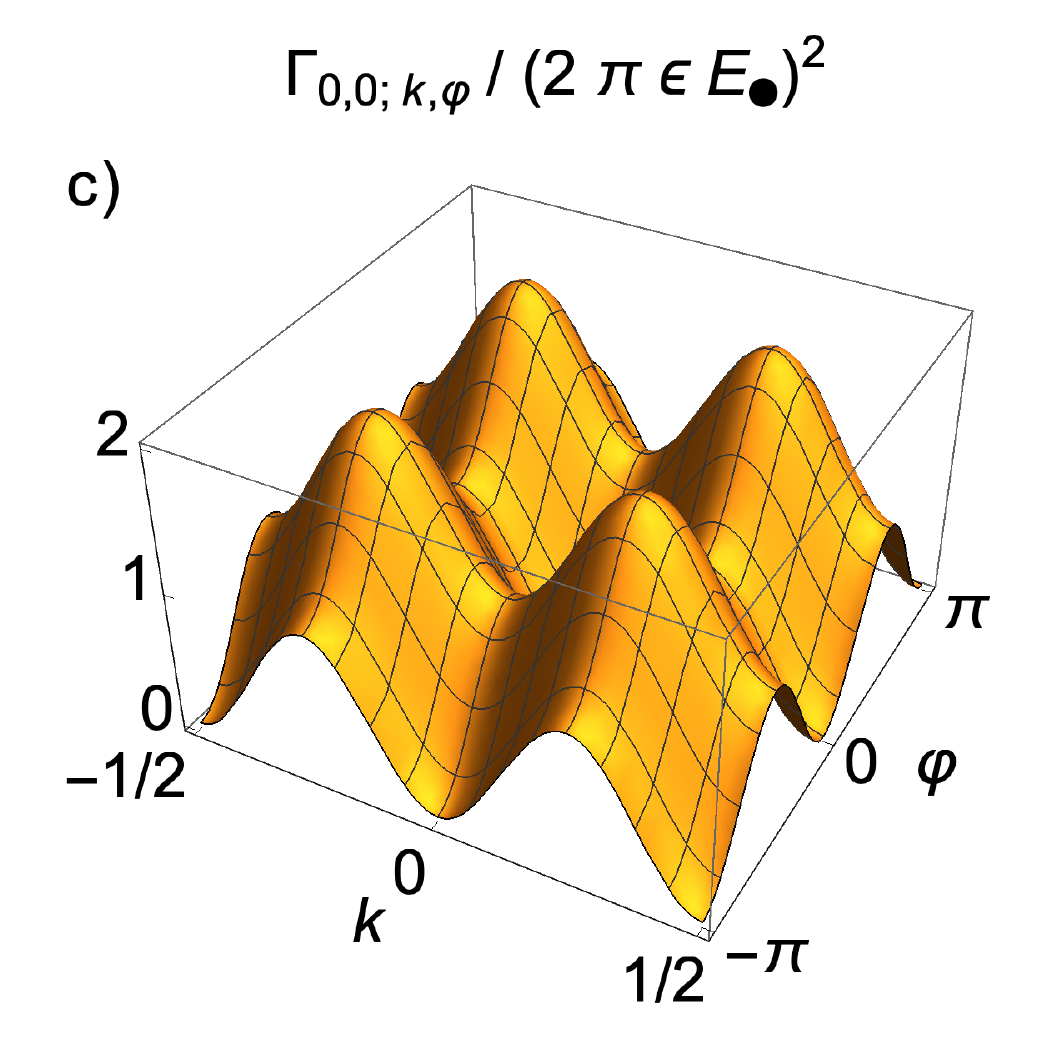}  
      \includegraphics[height=4.27 cm]{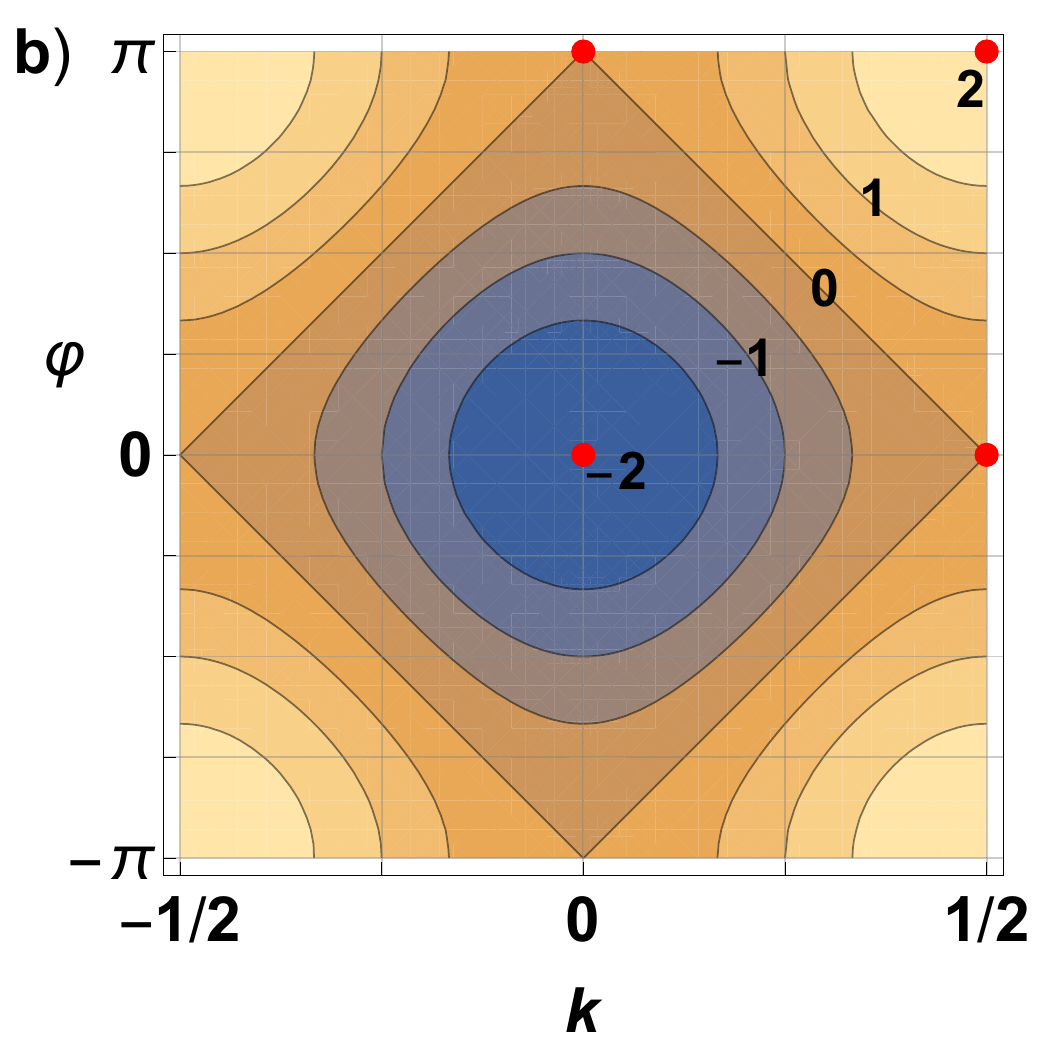}
\includegraphics[height=4.27 cm]{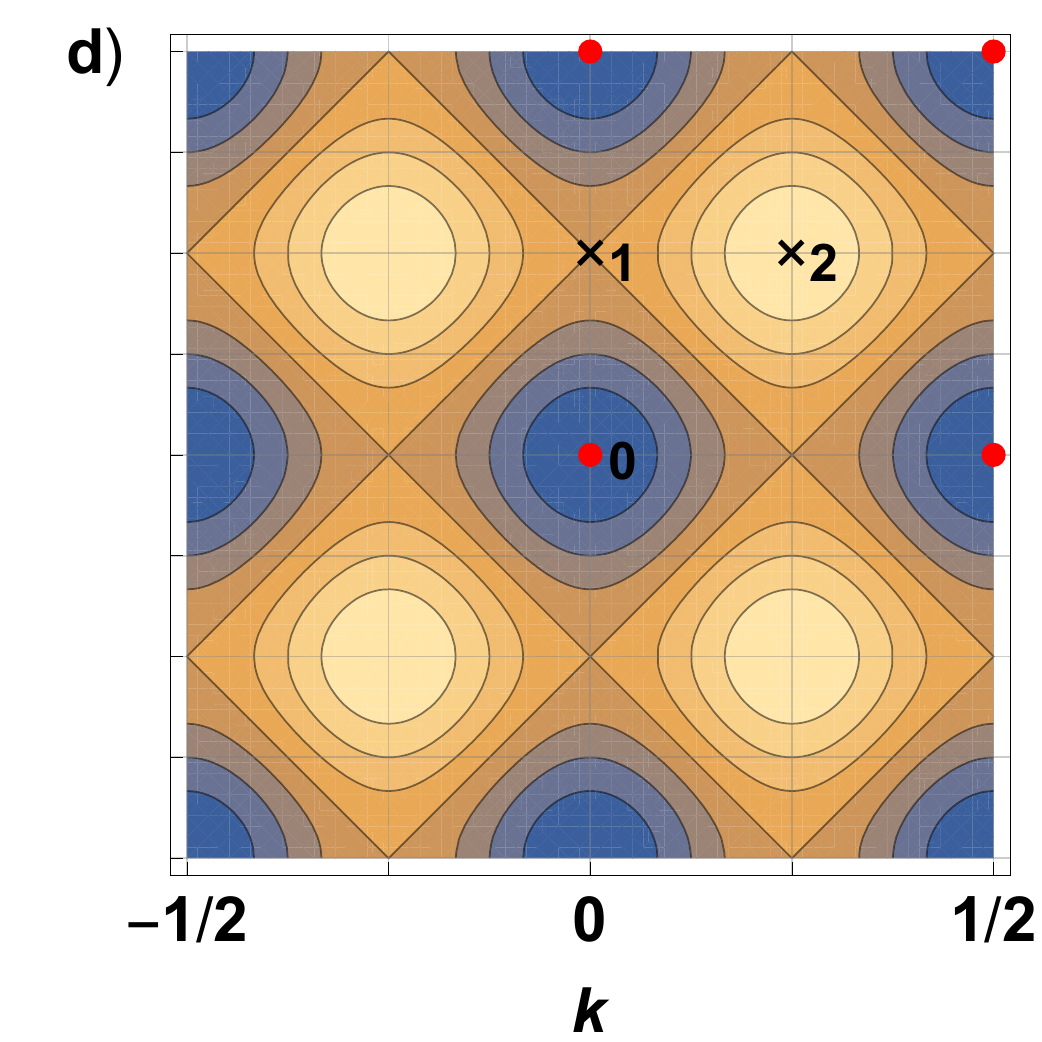}

\caption{(a) and (b)  Three-dimensional and contour plots of the elementary dualmon circuit energy spectrum, $E_{k,\varphi}$,  \cref{Eq:IdealEnergies}. The four critical points, where the gradient vanishes, are indicated by red circles. The saddle points are only degenerate if $E_Q=E_J$, but the locations of the critical points are fixed.     (c) and (d) The pure dephasing rate $\Gamma_{0,0;\,k\!,\varphi}$ of superpositions of the  dualmon ground state $\ket{0,0}$ with other eigenstates $\ket{k,\varphi}$, \cref{eqn:Gamma}, due to fluctuations in both external bias charge and flux.   The dephasing rate vanishes  at the  critical points \mbox{$\ket{k,\varphi}\in\{{\ket{0,\pi}},{\ket{1/2,0}},{\ket{1/2,\pi}}\}$}. For illustrative purpose, we have chosen $E_Q=E_J\equiv E_\bullet$ and uniform noise $\epsilon_{n}=\epsilon_{\phi}\equiv \epsilon$. Selected function contour and extremal values are marked. 
    }
    \label{Fig:IdealEnergySpectrum}

\end{figure}

The eigenenergies of $\hat H$ are 
\begin{equation}
E_{k,\varphi}= -E_Q \cos ( 2\pi k ) - E_J \cos ({\varphi}). \label{Eq:IdealEnergies}
\end{equation} 
This spectrum, as shown in Figs. \ref{Fig:IdealEnergySpectrum}a and \ref{Fig:IdealEnergySpectrum}b, has four critical eigenstates 
$$\{ \ket{0,0}, \ket{0,\pi}, \ket{1/2,0}, \ket{1/2,\pi} \},$$
where $\nabla E_{k,\varphi} = (\partial_k E_{k,\varphi}, \partial_{\varphi} E_{k,\varphi}) = 0$, which are the ground state, two saddle points, and the maximally excited state respectively. Notably, the saddle points can be made degenerate when $E_Q = E_J$. We will show that the sensitivity to charge and flux noise vanishes at these critical points to linear order. \\

Expanding the eigenstates in the phase or number bases
\begin{equation}
    \left|k,\! \varphi \right\rangle \!= \!\!\!\!\sum\limits_{j = -\infty}^{\infty} \!\!\!e^{i 2\pi j k }\!\ket{\varphi \!-\! 2\pi j }_{\!\bar\phi}
    =\!\tfrac{e^{i k \varphi}}{\sqrt{2\pi}}\!\!\sum\limits_{j=-\infty}^{\infty} \!\!\!e^{-i j \varphi } \!\ket{ j\!-\!k }_{\!\bar n}\!,     \label{eq:Zakbasis}
\end{equation}
makes it apparent that the GKP codewords are eigenstates of the circuit. Particularly, following the definitions from Ref. \cite{Gottesman01} we can see \mbox{$\ket{\bar0_{\rm GKP}}\!=\!\ket{0,0}$} and \mbox{$\ket{\bar1_{\rm GKP}}\!=\!\ket{0,\pi}$}. The double-Bloch eigenstates in \cref{eq:Zakbasis} satisfy the normalisation \mbox{$\langle k,\!\varphi\ket{k'\!,\!\varphi'}\!=\!\delta(k\!-\!k')\delta(\varphi\!-\!\varphi')$}, {and}
the generalised periodic boundary identities
\mbox{$ \ket{ -1/2, \varphi} \! =\! \ket{ 1/2, \varphi }$} and \mbox{$\ket{ k, - \pi} \!=\! e^{-i 2\pi k} \!\ket{ k, \pi}$}.

\subsection{Perturbative charge and flux noise in the elementary circuit}

We include the effect of external voltage and flux noise, as shown in \cref{Fig:circuit}b, with  an external gate voltage $V_x(t)$ coupled capacitively via a parasitic capacitor $C_x$, and an external flux $\Phi_x(t)$  through the circuit loop \cite{Devoret97}. In \cref{appen:derivationHamil}, we show that for small $C_x$ the Hamiltonian of the circuit becomes
\begin{equation}
\hat H'(t)\!=\! - E_Q \cos \!\big(2\pi\! \left(\hat{n}\!+\! n_x(t)\right)\!\big) - E_J \cos\!\big(\hat{\phi} \!+\!  \phi_x(t) \big), \label{Eq:Ideal_DCNoise_Hamiltonian}
\end{equation}
where \mbox{$n_x(t) = {C_x V_x(t)}/{(2e)}$} and \mbox{$\phi_x(t) = 2\pi \Phi_x(t)/{\Phi_0}$}.

For fixed values of the biases, the eigenenergies are \mbox{$\bar E_{k,\varphi}(n_x,\phi_x)\!= \!-E_Q \cos ( 2\pi (k+n_x) )\! -\! E_J \cos ({\varphi}\!+\!\phi_x)$}.  In linear response, the sensitivity of the system to small variation in bias is therefore given by 
\begin{equation}
\big(\partial_{n_x} \bar E_{k,\varphi},\partial_{\phi_x} \bar E_{k,\varphi}\big)\big|_{n_x=\phi_x=0}=\nabla E_{k,\varphi}.
\end{equation}
For  time-dependent noise in the bias parameters, this result implies that the noise sensitivity is determined by  $\nabla E_{k,\varphi}$, so that the  noise sensitivity vanishes  at the critical points, at linear order.

For small-amplitude noise, we expand  the Hamiltonian to linear order in the noise terms, so
\begin{equation}
    \hat{ H }'(t) \simeq \hat{ H } + n_x(t) \hat{ {A}}_{n} + \phi_x(t) \hat{{A}}_{ \phi},
\end{equation}
where
\begin{subequations}
\begin{eqnarray}
    \hat{{A}}_{n} &=& 2\pi E_Q \sin (2\pi \hat{n}), \label{eq:An} \\
         \hat{{A}}_{ \phi} &=& E_J \sin(\hat{\phi}). \label{eq:Aphi}
\end{eqnarray}
\end{subequations}
Since $[\hat H,\hat A_{n}]= [\hat H, \hat A_{\phi}]=0$, flux and charge noise do not induce transitions between eigenstates.  Further, the critical states are null vectors of $\hat A_{n,\phi}$, so they are immune to charge and flux noise {to first order}.

We illustrate the general noise insensitivity using the example of uncorrelated charge and flux noise, for which
\begin{subequations}
\begin{eqnarray}
    \left\langle n_x(t) n_x(t') \right\rangle &=& \epsilon_{n}^2 \delta (t-t'),  \\
\left\langle \phi_x(t) \phi_x(t') \right\rangle &=& (2\pi  \epsilon_{\phi})^2 \delta (t-t'), \\
\left\langle n_x(t) \phi_x(t') \right\rangle &=& 0,
\end{eqnarray}
\end{subequations}
where $\epsilon_{n}$ and $\epsilon_{\phi} $ are noise amplitudes. 
With this white noise model, the  evolution of the system density matrix, $\rho$, is given by the master equation  \cite{Budini01,Stace02}
\begin{equation}
\dot\rho(t) = - \tfrac{i}{\hbar} [ \hat{ H }, \rho(t) ] + 2\epsilon_{n}^2 \mathcal{D} [\hat{{A}}_{n}] \rho (t) 
+ 2 (2\pi \epsilon_{\phi})^2 \mathcal{D}[\hat{{A}}_{\phi}] \rho(t), 
\end{equation}
where  $    \mathcal{D}[A] \rho =  A\rho A^\dag - (\rho A^\dag A + A^\dag A \rho )/2$.

To calculate the decoherence rate between superpositions of eigenstates, we
 suppose that  the system is initially in a pure state \mbox{$ \ket{\psi} \!=\! \mu\! \ket{ k,\varphi }\!+\! \mu'\! \ket{ k'\!,\varphi' }$}, so that \mbox{$\rho(0) \!=\! {\ket{\psi}}{\bra{\psi}}$}. 
  Off diagonal elements are  right eigenoperators of the lindblad superoperators, but generally with non-zero eigenvalues (i.e.,\ dephasing rates), 
 \begin{align}
 \mathcal{D} [\hat{{A}}_n]\ket{k,\!\varphi}\!\bra{k'\!,\!\varphi'}={}&\! -\tfrac12(2\pi E_Q)^2 \gamma_{2\pi k,2\pi k'}\ket{k,\!\varphi}\!\bra{k'\!,\!\varphi'},\nonumber\\
  \mathcal{D} [\hat{{A}}_\phi]\ket{k,\!\varphi}\!\bra{k'\!,\!\varphi'}={}& - \tfrac12E_J^2 \gamma_{\varphi,\varphi'}\ket{k,\!\varphi}\!\bra{k'\!,\!\varphi'},\nonumber
 \end{align}
 where 
 \begin{equation}
     \gamma_{y,y'}\!=\!(\sin (y) \!-\! \sin (y'))^2. \label{eq:function_gamma}
 \end{equation}
 The pure dephasing rate is therefore given by 
 \begin{equation}
 \Gamma_{k,\varphi;\,k'\!,\varphi'}=(2\pi \epsilon_n E_Q )^2 \gamma_{2\pi k,2\pi k'}+ (2\pi  \epsilon_\phi E_J)^2 \gamma_{\varphi,\varphi'}.\label{eqn:Gamma}
 \end{equation}
For any choice of $k$, there are values of $k'$ for which $\gamma_{2\pi k,2\pi k'}\!=\!0$, and similarly for \mbox{$\varphi$ and $\varphi'$}. Of particular interest is the fact that for superpositions of the  critical eigenstates both lindblad superoperators vanish, $\gamma_{2\pi k,2\pi k'}\!=\!\gamma_{\varphi,\varphi'}=0$, so that \mbox{$ \Gamma_{k,\varphi;k'\!,\varphi'}\!=\!0$} when   \mbox{${\ket{k,\varphi}}, {\ket{k'\!,\varphi'}}\!\in\!\{{\ket{0,0}},{\ket{0,\pi}},{\ket{1/2,0}},{\ket{1/2,\pi}}\}$}.  We plot \mbox{$ \Gamma_{0,0;k,\varphi}$} in Figs. \ref{Fig:IdealEnergySpectrum}c and \ref{Fig:IdealEnergySpectrum}d, showing that the dephasing rates from fluctuations in both external bias charge and flux vanish at the critical points.
This property makes {these} states intriguing candidates for robustly storing quantum information. 

\section{ Realistic circuit elements} \label{sec:parasitic}

 Realistically, the dualmon circuit will have some linear inductance $L$ in the ring, and capacitance $C$ across the JJ \cite{Matveev02}. We therefore extend the model to account for the effects of these parasitic elements, and we show that the noise insensitivity of the elementary circuit is retained under certain assumptions for the circuit parameters. 
 
 The resulting lumped-element model, shown in \cref{Fig:circuit}c,  has an additional circuit node, and the Hamiltonian for the realistic circuit is
\begin{eqnarray}
\hat{ \mathcal{H} }_{\mathrm {sys}}  &=&  E_C \hat{n}_A^2 + E_L ( \hat{\phi}_A - \hat{\phi}_B  )^2 \nonumber \\
&& - E_Q \cos \left( 2\pi \hat{n}_B \right) - E_J \cos (\hat{\phi}_A),
\end{eqnarray}
where $E_C = {(2e)^2}/{(2C)}$, $ E_L ={\Phi_0^2}/{(8\pi^2L)}$, and the modes labeled $A$ and $B$ refer to the circuit nodes indicated in \cref{Fig:circuit}c. We note that models of this form have been studied in Ref. \cite{Ganeshan16}.

\subsection{Energy bands} \label{subsec:bands}

We assume that $L$ and  $C$ are small, so that \mbox{$E_L, E_C \gg E_Q, E_J$}. 
In this case, the high-frequency {dynamics of the associated} $LC$ oscillator will dominate, so it is convenient to
 transform to new conjugate coordinates
\begin{equation}
\hat{\phi}_1 \!  =  \!  \hat{\phi}_A  \!  -  \!  \hat{\phi}_B,\hspace{0.1cm}  \hat{n}_1  \!  =  \! \hat{n}_A, \hspace{0.3cm} \hat{\phi}_2  \!  =  \!  \hat{\phi}_B,\hspace{0.1cm}  \hat{n}_2  \!  =  \!  \hat{n}_A  \!  +  \!  \hat{n}_B.  \label{coords}
\end{equation}
In these coordinates we have 
\begin{equation}
\hat{ \mathcal{H} }_{\mathrm{sys}} = {\hat{  H }_{\mathrm{HO}} + \hat{\mathcal V}},\label{Eq:NonIdeal_Hamiltonian_Mode12}
\end{equation}
where
\begin{eqnarray}
    \hat{  H }_{\mathrm{HO}} &=& E_C \hat{n}_1^2 + E_L \hat{\phi}_1^2, \\
    \hat{\mathcal V} &=& - E_Q \cos ( 2\pi ( \hat{n}_1 \!-\! \hat{n}_2 ) )\!-\! E_J \cos ( \hat{\phi}_1 \!+\! \hat{\phi}_2 ). \quad \quad
\end{eqnarray}
$\hat{ \mathcal{H} }_{\mathrm{sys}}$ commutes with $\cos (2\pi \hat n_2)$ and $\cos (\hat \phi_2)$, so eigenstates of $\hat{ \mathcal{H} }_{\mathrm{sys}}$ will be  simultaneous eigenstates of these operators, which  are the Zak basis states $\ket{k,\varphi}_2$ \footnote{Here the subscript $i=1,2$ indicates states associated to mode $i$.}.  Thus, the eigenstates of $\hat{ \mathcal{H} }_{\mathrm{sys}}$ take the form
\begin{equation}
    \ket{\Psi_{m;k,\varphi}}_{1,2}=\ket{\psi_m(k,\varphi)}_1\ket{k,\varphi}_2, \label{eq:statemode12}
\end{equation}
 where $\ket{\psi_m(k,\varphi)}_1$ are the eigenstates  of the reduced Hamiltonian acting on \mbox{mode 1},
\begin{equation}
\hat{ H }_1({k,\varphi}) =\hat{  H }_{\mathrm{HO}}  + \hat{V}({k,\varphi}), \label{Eq:NonIdeal_HamiltonianMode1}
\end{equation}
with
\begin{equation}
    \hat{V}(k,\varphi)\!=\!- E_Q \cos (2\pi (\hat{n}_1\! - k)) \!-\! E_J \cos (\hat{\phi}_1\! + \varphi).
\end{equation}
The eigenenergies $E_{m;k,\varphi}$ of the two-mode Hamiltonian $\hat{\mathcal H}_{\mathrm{sys}}$, markedly, coincide with those of the mode-1 Hamiltonian $\hat{H}_1(k,\varphi)$. 

We are concerned with the limit $ E_L, E_C \gg E_Q, E_J$, in which $\hat{ H }_1({k,\varphi})$ describes a weakly nonlinear oscillator (i.e.,\ mode 1) that depends parametrically on the quantum numbers $k$ and $\varphi$ associated to \mbox{mode 2}.
We treat $\hat{V}(k,\varphi)$ perturbatively, and so we denote the eigenstates of $\hat{  H }_{\mathrm{HO}}$ as  $\ket{\psi_m^{(0)}}_1$ with eigenenergies \mbox{$E_{m}^{(0)} =(m+1/2)\hbar\Omega$}, where
\begin{equation}
\hbar\Omega=2 \sqrt{E_C E_L}.
\end{equation}
Within the oscillator ground state manifold, $m=0$, the first order perturbative correction  to the energy is 
\begin{eqnarray}
     E_{m=0;k,\varphi}^{(1)} &=& \tensor[_{1}]{\langle \psi_0^{(0)} | \hat V (k,\varphi)| \psi_0^{(0)} \rangle}{_{1}} \nonumber \\
&=& -E'_Q\cos(2\pi k)-E'_J \cos(\varphi), \label{pertenergy}
\end{eqnarray}
where $E'_Q=e^{- \pi^2 /z} E_Q$ and $E'_{J}=e^{- z /4 }E_J$ are renormalised QPS and JJ parameters arising from zero-point motion of mode 1, and $z=\sqrt{E_C/E_L}=\sqrt{L/C} \,(2e)^2/\hbar$ 
 is a dimensionless oscillator impedance. 
Equation \eqref{pertenergy} shows that within the oscillator ground state manifold,  mode 2 is governed by the elementary Hamiltonian $\hat H$ in \cref{Eq:Ideal_Hamiltonian}, with renormalised QPS and JJ energies.  Significantly, the critical points remain at the same locations in the double-Bloch band.

\begin{figure}[t]
\includegraphics[height=4.27cm]{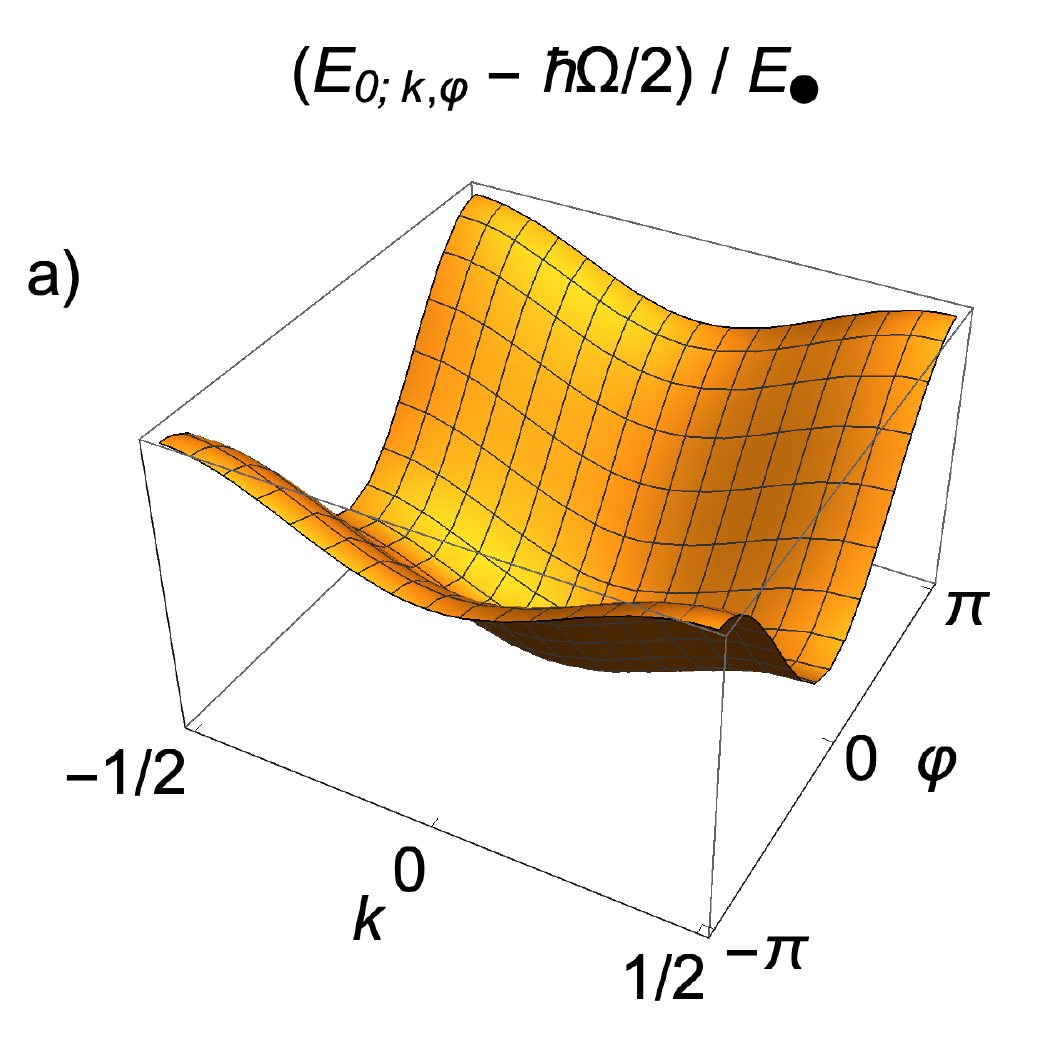} 
\includegraphics[height=4.27cm]{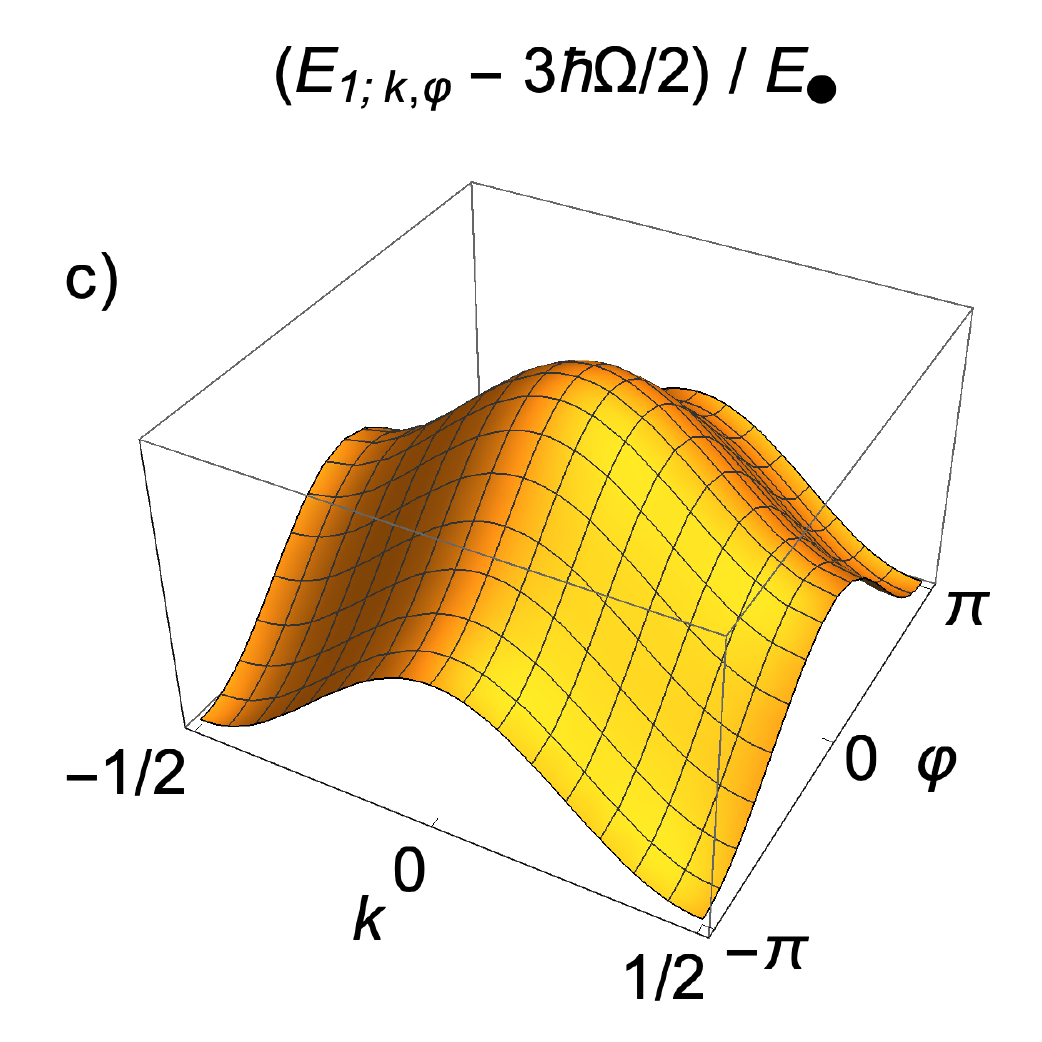} 
\includegraphics[height=4.27cm]{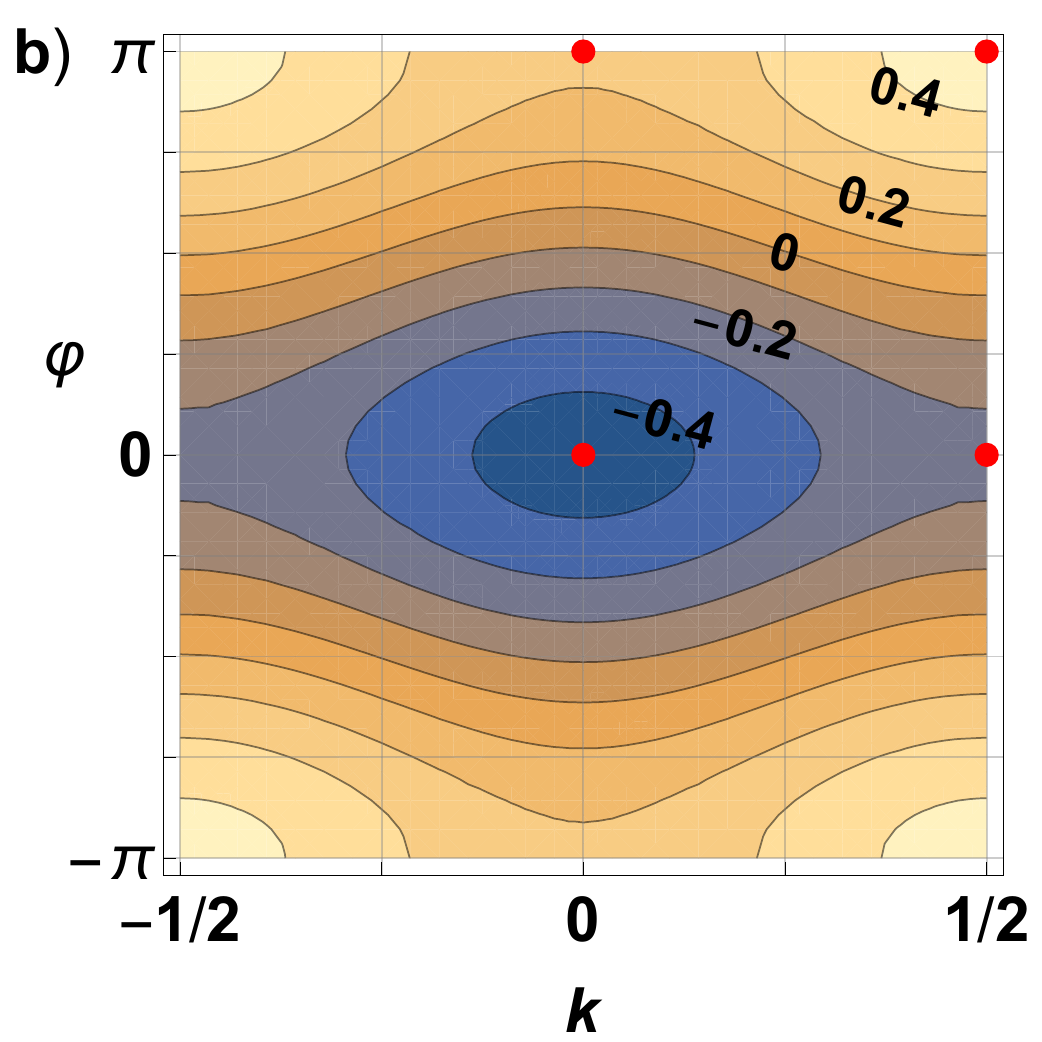} 
\includegraphics[height=4.27cm]{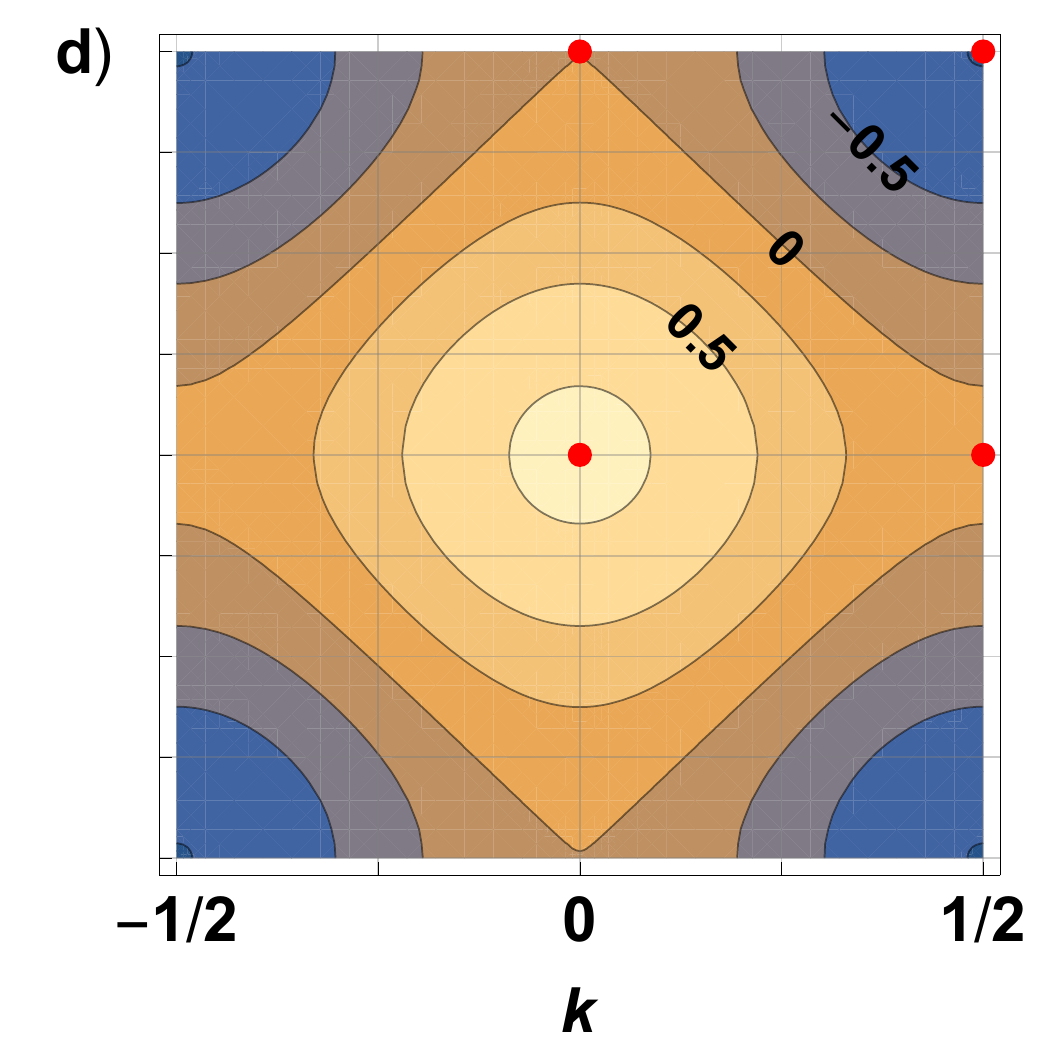} 
\caption{
(a) and (b) The ground state band energy, $E_{m=0;k,\varphi}$, of the realistic dualmon circuit relative to the unperturbed harmonic oscillator ground state energy, $E_{m=0}^{(0)}$. (c) and (d) The first excited state band energy, $E_{m=1;k,\varphi}$. 
Plots are generated by numerically solving the eigenvalue problem (see Appendix \ref{appen:eigenproblem}).  
On this scale,  the difference between the numerical values and the perturbative result in \cref{subsec:bands} is less than $0.005$ across the band.
For illustrative purposes, plots are drawn with $E_Q=E_J\equiv E_\bullet$, $E_C/E_J=200$ and \mbox{$E_L/E_J=10$}.  For these parameter values, $z=\sqrt{20}$, $E_Q' =0.11 E_\bullet$, $E_J'=0.33E_\bullet$, $E_Q'' = -0.38E_\bullet$, and $E_J''=-0.40E_\bullet$.}
\label{Fig:NonIdeal_BandStructure}
\end{figure}

 To verify the perturbative arguments above, we  numerically solve for  $\ket{\psi_m(k,\varphi)}_1$ and eigenenergies $E_{m;k,\varphi}$ {non-perturbatively} in the Zak basis for mode 1, as described in \cref{appen:eigenproblem}. 
 Figures \ref{Fig:NonIdeal_BandStructure}a and \ref{Fig:NonIdeal_BandStructure}b show the oscillator ground state manifold  energy band $E_{m=0;k,\varphi}$ relative to the unperturbed harmonic oscillator ground state energy,  for  $E_L,E_C\gg E_Q=E_J$. 
 The nonlinear energies $E_{J,Q}$  are renormalised to $E'_{J,Q}$, but the band structure is otherwise qualitatively the same as  the elementary case, and with the same critical eigenstates.  We show below that the noise sensitivity vanishes at the critical eigenstates, as for the elementary circuit.
 
 Figures \ref{Fig:NonIdeal_BandStructure}c and \ref{Fig:NonIdeal_BandStructure}d  also demonstrate the first excited manifold, $E_{m=1;k,\varphi}$, relative to the unperturbed harmonic oscillator first excited state energy.  In this manifold, the renormalised nonlinear energies are \mbox{$E''_Q\!=\!(1\!-\!2\pi^2/z)E'_Q$} and \mbox{$E''_J\!=\!(1\!-\!z/2)E'_J$}, so that the first excited band may be inverted relative to the ground state manifold (i.e.,\ the locations of minima and maxima are exchanged).  It follows that the interband transition energies at the critical points can be made non-degenerate, facilitating spectroscopic addressability of  the critical states. 
This addressability provides an avenue to state preparation: spectroscopic measurements of a given accuracy will localise the system in a narrow range of $k$ and $\varphi$ near the observed transition energy. This will be discussed in more details in \cref{sec:waveguide}.
 
\subsection{Perturbative charge and flux noise in the realistic circuit} \label{subsec:classicalnoise}

 We introduce charge and flux noise in the same manner as for the elementary circuit \cite{You19}. The Hamiltonian is then 
\begin{eqnarray}
    \hat{ \mathcal{H} }'_{\mathrm{sys}} &=& E_{C_\Sigma} ( \hat{n}_A + n_{x}(t) )^2 \!+\! E_L ( \hat{\phi}_A - \hat{\phi}_B -  \phi_{x}(t) )^2 \quad \quad \nonumber \\
&&- E_Q \cos ( 2\pi  \hat{n}_B  ) - E_J \cos ( \hat{\phi}_A  ), \label{Eq:NonIdeal_DCNoise_Hamiltonian}
\end{eqnarray}
where $E_{C_\Sigma} = {(2e)^2}/{(2C_\Sigma)}$ with $C_\Sigma=C+C_x$, and $n_x$ and $\phi_x$ are external bias terms, as  discussed earlier. In what follows, we assume that $C_x\ll C$, and take $E_{C_\Sigma}=E_C$ for simplicity. 

Constant charge and flux bias can be transformed away by suitable gauge choice \cite{koch09,Manucharyan09}, so we again consider the effect of zero-mean, white noise.
As before, we make the coordinate transformation given by \cref{coords}, and expand the Hamiltonian to linear order in the noise terms, so
\begin{equation}
\hat{ \mathcal{H} }'_{\mathrm{sys}} \simeq
\hat{ \mathcal{H} }_{\mathrm{sys}}
+n_x(t) \hat{ \mathcal{A}_n }
+\phi_x(t) \hat{ \mathcal{A}_\phi }, \label{eq:H_sys_smallnoise}
\end{equation}
where 
\begin{subequations}
\begin{eqnarray}
     \hat{\mathcal{A}}_{n} &=& 2E_{C} \hat{n}_1, \label{eq:noiseoperator_An} \\
     \hat{\mathcal{A}}_{\phi} &=& -2E_L \hat{\phi}_1.
\end{eqnarray}
\end{subequations}
The  master equation for the noisy system is then
\begin{equation}
\dot\varrho = - \tfrac{i}{\hbar} [ \hat{\mathcal H}_{\mathrm{sys}}, \varrho ] + 2 \epsilon^2_{n} \mathcal{D}[\hat{\mathcal{A}}_n] \varrho
 + 2 (2\pi \epsilon_{\phi})^2 \mathcal{D}[\hat{\mathcal{A}}_{\phi}] \varrho,  \label{eq:externalnoise_me}
\end{equation}
 where $\varrho$ is the {joint} density operator for modes $1$ and $2$.

Since $\hat{\mathcal{A}}_n$ and $\hat{\mathcal A}_{\phi}$ have action on the Hilbert space associated with mode 1 only, they are diagonal within any oscillator manifolds. Using the first order perturbative correction to the oscillator modes, we compute matrix elements of these dissipators in the ground state manifold (see \cref{appen:matrixelement} for more detail), and find that
\begin{subequations}
\begin{eqnarray}
     \!\tensor[_{1\!,2}]{\bra{\Psi_{0;k\varphi}} \!{\hat {\mathcal A}_n}\! \ket{\Psi_{0;k'\varphi'}\!} }{_{\!1\!,2}} \! &=& \!
   2\pi E'_Q \! \sin (2\pi k)\delta(k\!-\!k')\delta(\varphi \!-\!\varphi'), \label{eq:para_An} \nonumber \\ \\
    \!\tensor[_{1\!,2}]{\bra{\Psi_{0;k\varphi}} \!{\hat {\mathcal A}_{\phi}}\! \ket{\Psi_{0;k'\varphi'}\!} }{_{\!1\!,2}} \! &=& \!
    E'_J \!  \sin (\varphi)\delta(k\!-\!k')\delta(\varphi\!-\!\varphi'), \nonumber \\  \label{eq:para_Aphi}
\end{eqnarray}
\end{subequations}
which implies
\begin{subequations}
\begin{eqnarray}
     \Pi_0\hat{\mathcal{A}}_n\Pi_0 &=& 2 \pi E'_Q    \sin (2\pi \hat n_2), \label{eq:projectednoiseAn} \\
\Pi_0\hat{\mathcal{A}}_\phi \Pi_0&=&   E'_J    \sin ( \hat \phi_2), \label{eq:projectednoiseAphi}
\end{eqnarray}
\end{subequations}
where $\Pi_0 = \textstyle \int_{-1/2}^{1/2} dk \textstyle \int_{-\pi}^{\pi} d\varphi \ket{\Psi_{0;k,\varphi}}_{1,2} \bra{\Psi_{0;k,\varphi}}$ is the projection onto the ground state manifold.
Equations \eqref{eq:projectednoiseAn} and \eqref{eq:projectednoiseAphi} are of the same form as the dissipators for the elementary circuit, Eqs. \eqref{eq:An} and \eqref{eq:Aphi}, and so lead to dephasing rates of the same form.  That is, a superposition of two eigenstates in the ground state manifold, $\mu\ket{\Psi_{0;k,\varphi}}+\mu'\ket{\Psi_{0;k',\varphi'}}$,  
will dephase at a rate also given by  $\Gamma_{k,\varphi;\,k'\!,\varphi'}$ in  \cref{eqn:Gamma}, with the replacements $E_{Q,J}\rightarrow  E'_{Q,J}$.  
 As for the elementary circuit, the dephasing rate vanishes for superpositions of the critical states.

 \subsection{Parametric and non-perturbative noise sources in the realistic circuit
 } \label{subsec:parameter}

 The Aharonov-Casher interference \cite{Aharonov84} of phase slips, dependent on charge distribution, could be detrimental to QPS devices. Indeed, phase slips exhibit along  a strongly disordered superconducting nanowire \cite{Astafiev12}; interference of phase slips at different points may cause fluctuations in the phase slip energy $E_Q$ \cite{deGraaf18,Peltonen13}, and result in dephasing of coherent superpositions.  Phase slip interference on JJ chains (i.e., a granular model of superconducting nanowires \cite{Matveev02}) has also been observed experimentally \cite{Pop12}, and shown to be a decoherence source in the Josephson-junction-chain based fluxonium qubit \cite{Manucharyan12,Masluk12}. To mitigate such undesirable effect, a possible approach is to fabricate QPS nanowires in the form of short weak links \cite{Vanevic12,Peltonen16}, by which phase slips are limited to take place at the narrowest point, thus improving stability of the QPS energy.

The time-dependent charge noise considered  in \cref{subsec:classicalnoise} was perturbative, in which the standard-deviation in the charge was small, $\sigma_{n_x}\ll1/2$.  However, localised, two-level charge traps near the circuit can lead to a random-telegraph   charge fluctuation with $\sigma_{n_x}\sim1/2$, which is no longer perturbative.  This gives rise to quantum state diffusion, i.e., shifting of the system state in the charge space. However, as long as the noise magnitude is smaller than one electron worth of charge, we speculate that the errors could be correctable by means of active GKP quantum error correction \cite{Gottesman01}, for which GKP codes were originally designed. A short discussion on specific implementation of such active correction is included in \cref{sec:resonator}.

\section{Coupling to a waveguide} \label{sec:waveguide}

\begin{figure}[t]
    \centering
    \includegraphics[scale=0.5]{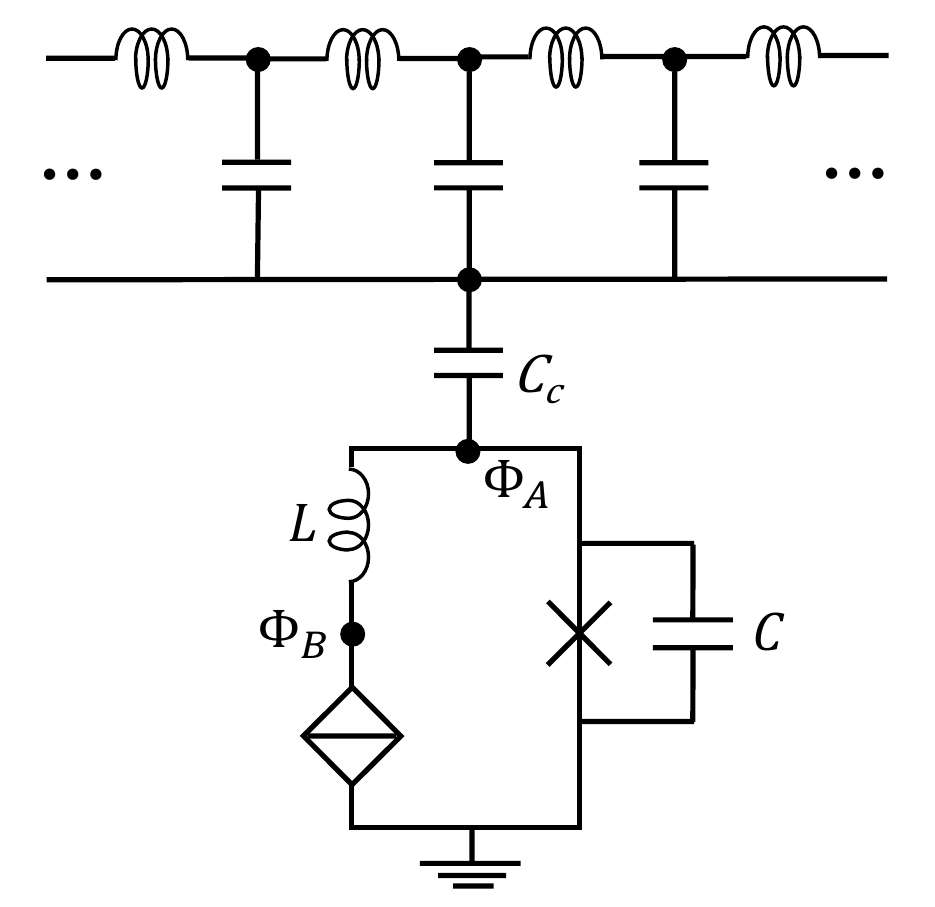}
    \caption{Lumped-element representation of the capacitive coupling between the realistic dualmon circuit and a waveguide.}
    \label{fig:nonideal_coupling}
\end{figure}

Realistic devices will need to couple to waveguides with which one can initialise, control, and perform readout on the system. Here we  couple the dualmon circuit capacitively to a waveguide and analyse spectroscopy on the device. We then propose a spectroscopic procedure for state preparation. 

The waveguide coupling also introduces quantum and thermal noise to the system, so we analyse the additional noise arising from this coupling. As for classical charge and flux noise, we show  that superpositions of the critical states are robust with respect to the induced quantum noise, for essentially the same reason, i.e., the band gradient vanishes at the critical points.

\subsection{Transmission spectrum} \label{sec:spectrum}

Figure \ref{fig:nonideal_coupling} illustrates a circuit where the dualmon is capacitively coupled to a waveguide. A detailed Hamiltonian derivation for this circuit is given in \cref{append:waveguide}. The Hamiltonian for the dualmon system coupled to the waveguide is
\begin{equation}
    \hat{\mathcal H}_{\mathrm {tot}} = \hat{\mathcal H}_{\mathrm {sys}} + \hat{\mathcal H}_{\mathrm{wg}} + \hat{\mathcal H}_{\mathrm{coup}}, \label{eq:wg_Htot}
\end{equation}
where $\hat{\mathcal H}_{\mathrm{sys}}$ is the dualmon Hamiltonian defined in \cref{Eq:NonIdeal_Hamiltonian_Mode12}, 
\begin{equation}
\hat{\mathcal H}_{\mathrm{wg}} = \int d \omega \hbar  \omega  \hat{a}^\dag(\omega) \hat a (\omega),
\end{equation}
is the waveguide Hamiltonian describing a continuum of modes, and
\begin{equation}
    \hat{\mathcal H}_{\mathrm{coup}} = \int d\omega g (\omega) (\hat a^\dag (\omega) + \hat a (\omega)) \hat n_1, \label{eq:wg_coupling}
\end{equation}
is the interaction with $g(\omega)$ the coupling strength.

We make use of the input-output formalism \cite{gardiner1985input,Combes17} to calculate the transmission spectrum; relevant calculations are presented in more details in \cref{append:inout}. In particular, the dynamics of the dualmon system coupled  to the waveguide and driven by a coherent field of amplitude $\alpha $ is governed  by
\begin{equation}
    \dot \varrho = - \tfrac{i}{\hbar} [ \hat{  \mathcal H}_{\mathrm {drive}}, \varrho] + \mathcal{D}[\hat b] \varrho, \label{eq:nonideal_me}
\end{equation}
where
\begin{eqnarray}
     \hat{  \mathcal H}_{ \mathrm {drive}} &=& \hat{\mathcal H}_{\mathrm {sys}} - \tfrac{i \hbar}{2} \sqrt{\gamma} (\alpha e^{-i \omega_D t} \hat n^{+}_{1} - \alpha e^{i \omega_D t} \hat n^{-}_1), \label{eq:wg_HSLH} \quad \quad \\
   \hat b &=& \sqrt{\gamma} \hat n^{-}_{1} + \alpha e^{-i \omega_D t} \mathbbm{1}.  \label{eq:inout}
\end{eqnarray}
Here $\hat n^{+}_{1} \, (\hat n^{-}_{1}) $ is the lower (upper) triangularised version of  $\hat n_1$, $\omega_D $ is the drive frequency, and $\sqrt{\gamma}$ is proportional to $g(\omega_D)$ determining bandwidth of the spectrum. The transmission is defined as
\begin{equation}
    T = \big | \langle \hat b \rangle/\alpha \big |^2, \label{eq:nonideal_transmission}
\end{equation}
 where
$
     \langle \hat b \rangle = \mathrm{Tr} \big (\hat b \varrho\big)$. 
 We solve the master equation (\ref{eq:nonideal_me}), use the obtained results to calculate $\langle \hat b \rangle$, and then get the transmission $T$. 
 
  \begin{figure}[t]
     \centering
     \includegraphics[width=0.9\columnwidth]{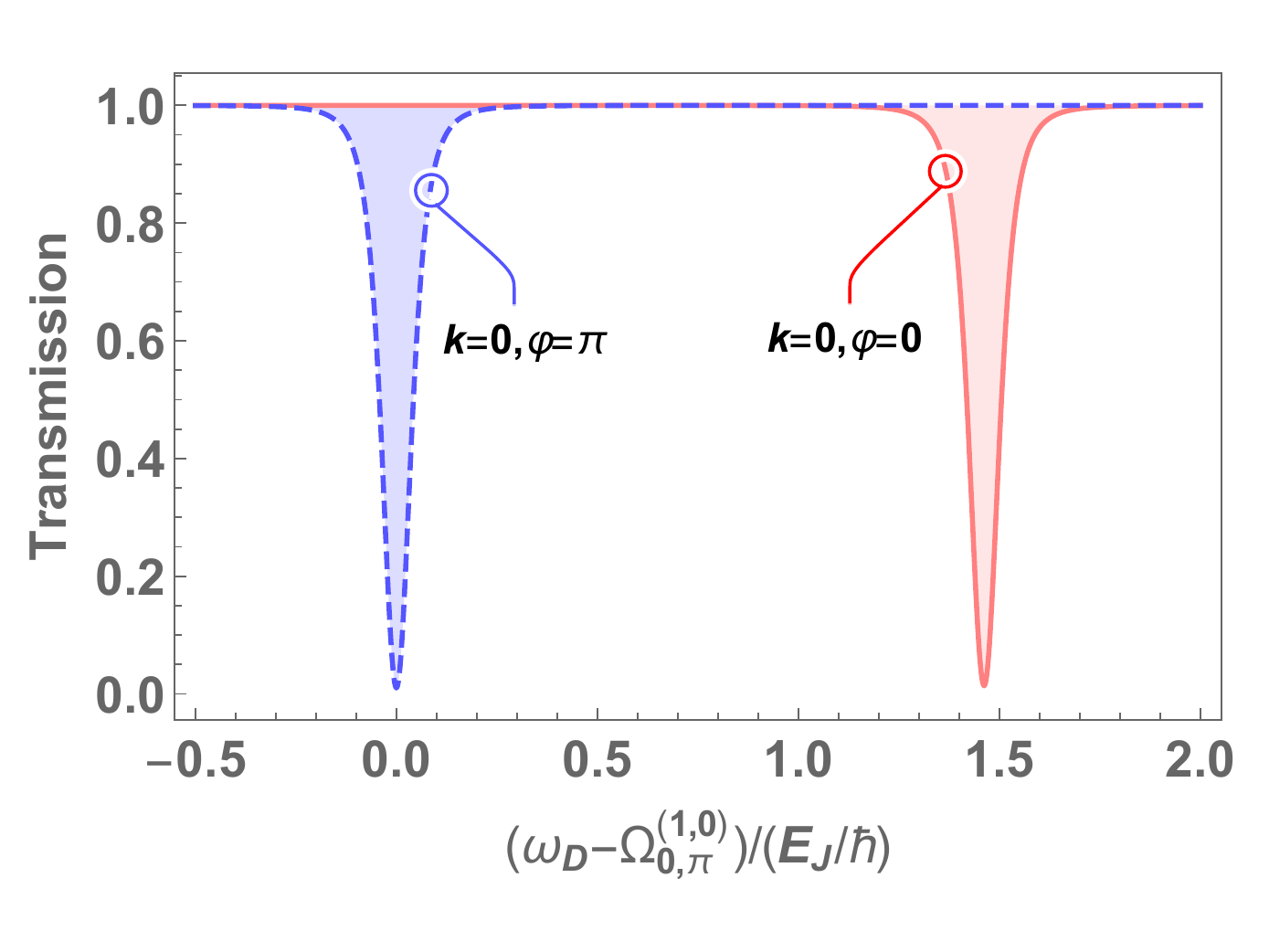}
     \caption{Transmission spectrum $T$ defined in \cref{eq:nonideal_transmission} with respect to the detuning of the drive frequency $\omega_D$ from the transition frequency $\Omega_{0,\pi}^{(1,0)}$ for two critical points, \mbox{$(k=0,\varphi=0)$ and $(k=0,\varphi=\pi)$}. Here $\Omega_{0,\pi}^{(1,0)}$ denotes the interband transition between the ground and first excited bands at the critical point $(k=0,\varphi=\pi)$. Relevant parameters for plotting are chosen as $E_Q =E_J,\hbar \Omega = 40\sqrt{5}\, E_J, z=\sqrt{20},$ $  \gamma =0.01\, \omega_D, $ and $\hbar \alpha^2 =0.1 \,E_J$.}      
     \label{fig:transmission}
 \end{figure}

The transmission spectrum assuming the circuit to be initialised in either one of two critical states $\ket{\Psi_{0;k=0,\varphi=0}}_{1,2}$ and $\ket{\Psi_{0;k=0,\varphi=\pi}}_{1,2}$ is displayed in \cref{fig:transmission}. For illustrative purpose, the transmission has been plotted in dependence on the detuning of the drive frequency $\omega_D$ from the transition between the lowest (i.e., ground) and first excited bands  for the critical point $(k=0,\varphi=\pi)$, denoted as $\Omega_{0,\pi}^{(1,0)}$ \footnote{We define
$\Omega_{k,\varphi}^{(m,n)} \equiv (E_{m;k,\varphi} - E_{n;k,\varphi})/\hbar$ as the transition frequency from the $n^{\mathrm{th}}$ band to the $m^{\mathrm{th}}$ band at a fixed pair of $(k,\varphi)$.}. It can be seen from \cref{fig:transmission} that resonance depends on the system state. This is consistent with the observation made in \cref{subsec:bands}, that is, the interband transition frequency is dependent on the state of the device, with the distinguishability of different $\ket{k,\varphi}$ states set by the difference in the interband transition frequency. 

\subsection{Qubit initialisation by spectroscopy} \label{subsec:stateprep}

The spectroscopic measurement described in the preceding subsection moreover gives a route to state preparation.
Since the interband transition frequency between the lowest and first excited bands of the dualmon is $(k,\varphi)$-dependent, transmission at a given frequency $\Omega_{k,\varphi}^{(1,0)}$ will localise the system around the corresponding value of $(k,\varphi)$ \footnote{Note that the mapping between interband frequencies $\Omega_{k,\varphi}^{(1,0)}$ and values of $(k,\varphi)$ is in general not unique, except for at the ground state $(k=0,\varphi=0)$, and the maximally exited state $(k=1/2,\varphi=\pi)$ (see \cref{Fig:transition} in \cref{append:inout}). Resolving states with similar transition frequencies can however be done by adjusting $n_x$ and $\phi_x$ while performing spectroscopy at a given frequency.}.
As shown in \cref{append:scan} scanning over transition frequency is equivalent to scanning over the two biases $n_x \in [-1/2,1/2)$ and $\phi_x \in [-\pi,\pi)$, one can thus localise the state to some value of $(k,\varphi)$
starting from a generic state such as a thermal state. Once the value of $(k,\varphi)$ has been localised, one can adjust $(n_x,\varphi_x)$ to move the origin of the band in~\cref{Fig:NonIdeal_BandStructure} such that the state becomes localised at any desired point, e.g., the ground state $(k=0,\varphi=0)$.

Choosing, e.g., $(k=0,\varphi=0)$ and $(k=0,\varphi=\pi)$ as the two operating points for the dualmon,
logical basis state preparation corresponds to preparing a state highly localised around one of these two values.
The exact form of the distributions over $(k,\varphi)$ is not particularly important, nor is it crucial that the states are pure, as long as they are sufficiently well-localised. This, perhaps counter intuitive, can be understood by making an analogy with approximate GKP codewords~\cite{Gottesman01}. Two approximate qubit codewords for the dualmon can be defined as
\begin{equation}
\ket{i_L} = \int dkd\varphi\,\psi_i(k,\varphi)\ket{k,\varphi},
\end{equation}
with $\psi_i(k,\varphi)$ some localised distribution around $(k=0,\varphi=0)$ for $i=0$ and $(k=0,\varphi=\pi)$ for $i=1$. A well-localised but mixed state can be viewed as an approximate qubit codeword with a set of small (approximately) correctable errors.
In other words, we can think of the well-localised mixed state as the codeword $\ket{i_L}$ sent through an error channel with the ``size'' of the errors bounded by the measurement resolution of the spectroscopic measurement.
This is thus analogous to noisy preparation of GKP codewords. In a universal scheme for quantum computing, which we will not discuss here, it is also crucial that the logical operations do not amplify errors too badly, i.e., turning small errors into large ones. Provided that this is true, the two distributions will remain well-localised throughout the computation, and can in principle be distinguished with high fidelity in a final measurement.

We have argued that it should be possible to prepare states well-localised around a single value of $(k,\varphi)$ using spectroscopic measurements, and furthermore such states can be thought of as imperfect, yet high-quality, qubit codewords. It is however a far more challenging task to prepare superposition states, such as $\ket{+_L} = \frac{1}{\sqrt 2}(\ket{0_L}+\ket{1_L})$, with a similar guarantee of ``small errors''.  Preliminary results suggest that introducing symmetry-breaking terms to the Hamiltonian, e.g., breaking the $2\pi$ periodicity of $\hat\phi_2$, can be used to prepare states of the form $\ket{+_L}$. The quality of such operations, as well as the possibility of universal control, will be the subject of future work.

\subsection{Pure dephasing from thermal and quantum noise in the waveguide}

The capacitive coupling to the waveguide, in addition to allow performing spectroscopy and state initialisation, opens a channel for noise to appear in the dualmon circuit. The induced noise is manifest via the coupling operator $\hat n_1$, which is diagonal within each manifold while coupling states among different manifolds. As we are interested in encoding the dualmon in the ground state manifold, say in a superposition of eigenstates $\mu\ket{\Psi_{0;k,\varphi}}_{1,2}+\mu'\ket{\Psi_{0;k',\varphi'}}_{1,2}$, the relevant decoherence source is either pure dephasing within the encoded manifold or transitions to other manifolds. The master equation for the dualmon density operator when considering ony pure dephasing in the ground state manifold is \cite{Breuer02} 
\begin{equation}
    \varrho_g \!=\! -\tfrac{i}{\hbar} [\hat{\mathcal H}_{\mathrm{sys},g}, \varrho_g] \!+\! 2\pi \! \lim_{\omega \to 0} \!  \big( J(\omega) (2 N(\omega) \!+\!1) \big)\mathcal D [\hat n_{1,g}] \varrho_g, \label{eq:me_groundmanifold}
\end{equation}
where $\varrho_g, \hat{\mathcal H}_{\mathrm{sys},g}$, and $\hat n_{1,g}$ are all projections onto the ground state manifold, $J(\omega) = 1/\hbar^2 \int d\omega' g^2 (\omega) \delta (\omega'- \omega) $ is the spectral density, and $N(\omega)$ is the thermal distribution at frequency $ \omega$. $\hat n_{1,g}$ can be straightforwardly computed from Eqs. \eqref{eq:noiseoperator_An} and \eqref{eq:projectednoiseAn}. 
The pure dephasing rate of the superposition of interest in the ground state manifold is then found to be
\begin{equation}
    \Gamma_{k;k'} =  \tfrac{ \pi^2 E'_Q}{E_C}   \lim_{\omega \to 0} \!  \big( J(\omega) (2 N(\omega) \!+\!1)\big)  \gamma_{2\pi k, 2\pi k'},
\end{equation}
where the function $\gamma_{y,y'}$ has been defined in \cref{eq:function_gamma}. In general this is not zero, due to thermal and vacuum fluctuations in the waveguide.  However, similar to  the case of classical noise in \cref{subsec:classicalnoise}, we have $\gamma_{2\pi k, 2\pi k'}=0$ for superpositions of the critical states, so that dephasing from waveguide fluctuations is  completely suppressed.

\subsection{Decoherence from thermal waveguide excitation} \label{subsec:thermal}

Another decoherence source stemming from the waveguide coupling is thermal transitions among different manifolds. We show below that provided the harmonic oscillator gap is much larger than $k_B T$ the thermal induced dephasing rate is very small. For this purpose, we express the coupling operator $\hat n_1$ in the form
\begin{equation}
    \hat n_1 = \int_{-1/2}^{1/2} dl \int_{-\pi}^{\pi}  d \theta \sum_{m,n=0}^{\infty} \hat n_{1;l,\theta}^{(m,n)}, \label{eq:n1_decom}
\end{equation}
where \mbox{$\hat n_{1;l,\theta}^{(m,n)} \!=\! \tensor[_{1,2}]{\left\langle \Psi_{m;l,\theta} | \hat n_1 | \Psi_{n;l,\theta} \right\rangle}{_{1,2}} \ket{\Psi_{m;l,\theta}}_{1,2}\! \bra{\Psi_{n;l,\theta}}$} are eigenoperators of $\hat{\mathcal H}_{\mathrm{sys}}$ satisfying the commutation relations $ \big[ \hat{\mathcal H}_{\mathrm{sys}}, \hat n_{1;l,\theta}^{(m,n)} \big] = \hbar \Omega_{l,\theta}^{(m,n)} \hat n_{1;l,\theta}^{(m,n)}$. The master equation for the dualmon density operator, that accounts for the transitions between manifolds, is then given by \cite{Breuer02}
\begin{eqnarray}  
     &&  \dot  \varrho = -\tfrac{i}{\hbar } [\hat{\mathcal H}_{\mathrm{sys}}, \varrho] \nonumber \\
      && \! + \! \iint \! dl d\theta \sum_{m<n} \! \pi J (\Omega_{l,\theta}^{(n,m)}) (N(\Omega_{l,\theta}^{(n,m)})\!+\!1) \mathcal{D} \big[\hat n_{1;l,\theta}^{(m,n)}\big] \varrho \nonumber \\
      && \! + \! \iint \! dl d\theta \sum_{m>n} \! \pi J (\Omega_{l,\theta}^{(m,n)}) N(\Omega_{l,\theta}^{(m,n)}) \mathcal{D} \big[ \hat n_{1;l,\theta}^{(m,n)} \big] \varrho. \quad \label{eq:wg_masterequation}
\end{eqnarray}
In Eq. \eqref{eq:wg_masterequation}, the second and third lines respectively represent relaxations to lower manifolds and excitations to higher manifolds; notably, the two quantum numbers $l$ and $\theta$ are conserved, implying the absence of transitions within each manifold.  As well known from the standard two-level-system encoding, either relaxation or excitation dephases coherent superpositions. However, as the dualmon is encoded in superpositions of two eigenstates in the ground state manifold and there are no transitions between these two states, 
the dephasing rate for the dualmon is determined by excitations out of the ground state manifold only.  
In \cref{append:thermaldephasing}, we include several calculations comparing dephasing rates of the ordinary two-level-system encoding and ground-state-manifold one. The relevant dephasing rate, from Eq. \eqref{eq:wg_masterequation}, is proportional to $J(\Omega_{l,\theta}^{(m>0,0)}) N(\Omega_{l,\theta}^{(m>0,0)})$. The waveguide (the bath) is assumed to be in thermal state, i.e., \mbox{$N(\omega) = 1/(\exp( \hbar  \omega/ k_B T)-1)$} and as shown in \cref{append:waveguide}  $J(\omega) = \nu \hbar \omega$ with $\nu$ a constant, so 
\begin{equation}
    J(\omega) N(\omega) = \nu \hbar \omega/ (\exp{(\hbar \omega/k_B T)}-1),
\end{equation}
 which is very small for large $\hbar \omega/k_B T$. Hence, thermal dephasing is very weak as long as $  \hbar \Omega_{k,\varphi}^{(1,0)} \gg k_B T$. This condition is inherently satisfied with superconducting circuits, since the working temperature is typically very low.

 \section{Comparisons with related superconducting qubits} \label{sec:compare}

 It is instructive to point out the differences between the proposed circuit and other superconducting qubit devices previously discussed in the literature, namely, the $0-\pi$ qubit \cite{kitaev2006protected,Brooks13,Dempster14,Groszkowski2018,Paolo19}, the fluxonium \cite{koch09,Manucharyan09}, and the Aharonov-Casher device \cite{Bell16}. Specifically, the $0-\pi$ qubit uses two  nearly degenerate eigenstates with nearly disjoint support in coordinate space \cite{Groszkowski2018}, which affords its noise-rejecting properties. Realisation of the $0-\pi$ qubit would require implementing the $\pi-$periodic Josephson junction \cite{kitaev2006protected,Brooks13,Gladchenko09,Doucot12,Bell14,Smith19}.  In comparison, the robustness of the critical states of the dualmon device arises from the fact that the noise operators associated to weak external charge and flux noise commute with the eigenstates suppressing decay, and the gradient of the energy landscape vanishes at the critical states suppressing dephasing. 
 
The lumped circuit of the imperfect fluxonium \cite{koch09,Manucharyan09}, where phase slip is included at its superinductor \cite{Manucharyan12,Masluk12}, is remarkably identical to the realistic dualmon circuit shown in \cref{Fig:circuit}c. However, the essential difference between the two devices lies at the energy scales. In particular, the fluxonium employs a superinductor  \cite{koch09,Manucharyan09,Hazard19} to make the inductive energy very small; QPS on the superinductor is treated as an undesired noisy source \cite{Manucharyan12}. The JJ present in the fluxonium circuit generally works in the regime of large $E_J/E_C$. These are in contrast to the dualmon energy scales, which ideally would be $E_L, E_C \gg E_Q, E_J$ (see \cref{subsec:bands}). The difference at energy scales then results in the differences in the physics of the two devices. Indeed, fluxon tunneling is the primary process in the fluxonium \cite{Manucharyan12}, whereas the dual QPS and JJ elements in the dualmon allow tunneling of both fluxon and Cooper pair. The fluxonium wavefunction in flux space is localised within the wells of the Josephson cosine potential and in charge space is well-localised within a range of a Cooper pair  \cite{Manucharyan09}. The dualmon, instead, is completely delocalised in both flux and charge spaces (see Zak states in \cref{eq:Zakbasis}). Finally, the Aharonov-Casher device proposed in Ref. \cite{Bell16} is realised basically as a single JJ shunted by a superinductor, and thus closely resembles the fluxonium design; therefore, the above arguments also hold when comparing such device to the dualmon circuit. Furthermore, in that device the Aharonov-Casher interference effect is ultilised to suppress single fluxon tunneling while making pair tunneling dominant. This is different from the dualmon circuit, where only single fluxon tunneling is considered.

  We comment briefly on plausible experimental parameters for the dualmon circuit.   Josephson and  QPS energies can be routinely engineered to be \mbox{$E_J\sim E_Q\sim h \times 5$ GHz} \cite{Bouchiat98,Peltonen13,deGraaf18,Constantino18}. Parasitic capacitance and self-inductances  of order \mbox{$C\sim1$ fF} and \mbox{$L\sim 10$ nH} are feasible \cite{Bouchiat98,Peltonen13,Mooij05} corresponding to parameters for the dualmon circuit of $z\sim3$ and  \mbox{$\hbar\Omega\sim h \times 50 \textrm{ GHz}\gg E_J,E_Q$}, which is well within the validity range of the approximations employed in this work.

\section{Conclusions} \label{sec:conclusion}
  
We have shown that the dualmon circuit, encoded in superpositions of the critical points, is resilient against both classical and quantum white noise. The  device nonetheless could suffer from other noise sources, including the charge-dependent Aharonov-Casher effect in QPS devices that lowers the phase slip energy and large temporal charge fluctuations causing diffusion of quantum states.  Possibly, the former could be reduced by designing QPS elements in a weak-link form \cite{Vanevic12,Peltonen16}, whilst the later might be dealt with via active error correction of the GKP code \cite{Gottesman01}.

Any pair of the critical points could be chosen to represent a qubit; the nearly-degenerate saddle points might be a convenient choice. As discussed qualitatively in \cref{subsec:stateprep}, state preparation could be performed by interband spectroscopic measurements. The quality with which we can prepare states near the critical points then depends on the spectral resolution of such a measurement.

Relative phase shifts between the critical states  may be achieved using tunable JJ and QPS circuit elements, or by using  high frequency drive pulses to access excited states of the oscillator mode.  Tuning of JJ and QPS energies leaves the critical states unchanged, thus retaining the protected working space of the qubit; introducing tunable elements, however, opens the qubit to new dephasing channels.  More general control requires additional symmetry breaking terms in the Hamiltonian.
For example, a shunt inductor  switched transiently across the JJ breaks the discrete charge translation symmetry of the Hamiltonian, so it will  couple the Zak eigenstates of the original circuit.
Detailed analysis of control of the state space for one and multiple devices will be the subject of future work.

We conclude that the dualmon circuit considered here  offers a promising avenue for robustly storing quantum information, worthy of further study.  In particular, it hosts several critical eigenstates, superpositions of which are insensitive to both charge and flux noise  at linear order.  
This result holds even when the circuit includes  parasitic capacitance and inductance. Interestingly, the critical states are physical embodiments of the codewords of the GKP error correcting code. Active fault-tolerant preparation of such states in harmonic oscillators is extremely hard, so the dualmon circuit could possibly  offers an  alternative path towards accessing these states.

\begin{acknowledgments}
This research was supported by the Australian Research Council Centre of Excellence for Engineered Quantum Systems (EQUS, CE170100009) and a Discovery Early Career Research Award (DE190100380) as well as the Swiss National Science Foundation through the NCCR Quantum Science and Technology. TMS acknowledges visitor support from the Pauli Center for Theoretical Studies, ETH Zurich. We thank J. Cole and V. V. Albert for useful discussions.
\end{acknowledgments}

\appendix

\section{Derivation of the biased Hamiltonian, $\hat H'(t)$} \label{appen:derivationHamil}

Here we derive the Hamiltonian $\hat H'(t)$ in \cref{Eq:Ideal_DCNoise_Hamiltonian}. Applying the spanning tree model in Ref.\ \cite{Devoret97} for the circuit in \cref{Fig:circuit}b, we  choose the fluxes across the QPS and JJ to be $\Phi_Q = \Phi$ and $\Phi_J = \Phi + \Phi_x$, respectively.
The Kirchoff current conservation law at the active node of the circuit is
\begin{equation}
    \dot Q_Q + \dot Q_{C_x} + \dot Q_J =0.
\end{equation}
From the constitutive laws for  circuit elements, we know that the charge that has flowed through the  QPS is
\begin{equation}
Q_Q (\dot \Phi)  \equiv  \tfrac{2e}{2\pi}  \arcsin ( \dot \Phi/V_c),
\end{equation}
 the charge across the bias capacitor is \mbox{$Q_{C_x} \!=\! C_x (\dot \Phi \!- \! V_x)$}, and the current through the Josephson junction is \mbox{$\dot Q_J \equiv I_J = I_c \sin (2\pi (\Phi+\Phi_x)/\Phi_0)$}.   We thus find the  equation of motion for the circuit is
\begin{equation}
    \frac{d }{d t}   \big( Q_Q (\dot \Phi) +C_x  (\dot \Phi -  V_x)  \big) + I_c \sin \! \big( \tfrac{2\pi}{\Phi_0} (\Phi + \Phi_x) \big) \!=\! 0. 
\end{equation}
We identify terms above with the Euler-Lagrange equation of motion, $ \tfrac{d}{dt} Q - \partial_\Phi L = 0$ where $Q=\partial_{ \dot \Phi} L$ is the charge conjugate to $\Phi$, which implies
\begin{eqnarray}
 Q={\partial L}/{\partial \dot \Phi} &=& Q_Q (\dot \Phi) + C_x  \dot \Phi-Q_x,\label{Eq:Ideal_Noise_Deri_Mo1} \\
 {\partial L }/{\partial \Phi} &=& -I_c \sin \! \big( \tfrac{2\pi}{\Phi_0}  (\Phi + \Phi_x) \big),
\end{eqnarray}
where $Q_x = C_x V_x $. The Lagrangian is then given by
\begin{eqnarray}
L &=&\dot{\Phi} \,Q_Q(\dot \Phi)  + \tfrac{2 e V_c}{2\pi} \cos \big( \tfrac{2\pi}{2e}  {Q_Q(\dot \Phi)} \big) + \tfrac{C_x}{2} ( \dot{\Phi} - V_x  )^2 \nonumber \\
&& + \tfrac{I_c \Phi_0}{2\pi} \cos \big ( \tfrac{2\pi}{\Phi_0}  (\Phi + \Phi_x) \big ).
\end{eqnarray}
The Hamiltonian, $H(\Phi, \dot \Phi) \! = \! \dot \Phi Q \! -  \! L$, is
\begin{equation}
H(\Phi, \dot \Phi) \! = \! \tfrac{C_x}{2} \dot{\Phi}^2 \! -\!  E_Q \!
\cos\!  \big( \! \tfrac{2\pi}{2e}  Q_Q(\dot \Phi) \! \big)  
\! - \!E_J \! \cos\!  \big( \!  \tfrac{2\pi}{\Phi_0} (\Phi + \Phi_x) \! \big ), \label{Eq:Ideal_Noise_Deri_HamRaw}
\end{equation}
where $E_Q \!=\! 2eV_c/(2\pi)$, $E_J \!=\! I_c \Phi_0/(2\pi)$.
To compute $H(\Phi, Q)$ we need to invert \eqn{Eq:Ideal_Noise_Deri_Mo1} to find  \mbox{$\dot{\Phi} = \dot{\Phi} (Q)$}. Implicitly,   we have
\begin{equation}
     \dot{\Phi} =  V_c \sin \big( \tfrac{2\pi}{2e} (Q -C_x \dot{\Phi}+Q_x) \big). \label{Eq:Ideal_Noise_Deri_dotPhi1}
\end{equation}
We cannot solve this  analytically for $\dot\Phi$, however if $C_x \dot \Phi \ll Q$ we expand the right hand-side of \cref{Eq:Ideal_Noise_Deri_dotPhi1} in  powers  of $C_x$, and solve for $\dot\Phi$.  We find
\begin{eqnarray}
    \dot{\Phi} \! &=& \!  V_c \sin  \big(  \tfrac{2\pi}{2e}(Q + Q_x) \big) \nonumber \\
   && \times   \Big( 1 \!-\!  \tfrac{2\pi }{2e}C_x V_c \cos\! \big (\tfrac{2\pi}{2e}  (Q \!+\! Q_x) \big )  \Big) + O(C_x^2) .  \quad \quad  \label{Eq:Ideal_Noise_Deri_Cor}
\end{eqnarray}
Substituting this result  into \cref{Eq:Ideal_Noise_Deri_HamRaw}, we obtain
\begin{eqnarray}
H &=&  \!  - E_Q \cos  \!  \big (    \tfrac{2\pi}{2e} (Q  \!  +  \!  Q_x)    \big)   \!  -  \!  E_J \cos  \!  \big(   \tfrac{2\pi}{\Phi_0   } (\Phi + \Phi_x) \big)  \nonumber \\
&&  \!  {}+ E_{C_x} \cos  \!  \big(    \tfrac{4\pi}{2e} (Q  \!  +  \!  Q_x) \big) + O(C_x^2),
\end{eqnarray}
 where \mbox{$E_{C_x} \!=\! C_x V_c^2/4$}. Quantising the charge and flux operators, and defining $\hat n=\hat Q/(2e)$ and $\hat \phi=2\pi\hat\Phi/\Phi_0$, the Hamiltonian operator is then
\begin{eqnarray}
 \hat H' &=& - E_Q \cos \left( 2\pi (\hat{n} + n_x) \right)  - E_J \cos (\hat{\phi} + \phi_x) \nonumber \\
 && {}+  E_{C_x} \cos \left( 4\pi (\hat{n} + n_x) \right),
\end{eqnarray}
where $n_x = {Q_x}/(2e)$ and $ \phi_x = 2\pi \Phi_x/\Phi_0$. If the external capacitance is sufficiently small then \mbox{$   E_{C_x} \ll E_Q, E_J$}.  Since \mbox{$\big[\hat H,\cos \big(4\pi (\hat{n}+ n_x)\big)\big]=0$}, this term does not  change the eigenstates, and preserves the critical points.  We  therefore take $E_{C_x} =0$ for simplicity, yielding \cref{Eq:Ideal_DCNoise_Hamiltonian}.

\section{Eigenvalue problem for mode 1} \label{appen:eigenproblem}

\begin{figure}[t]
\includegraphics[height=4.5cm]{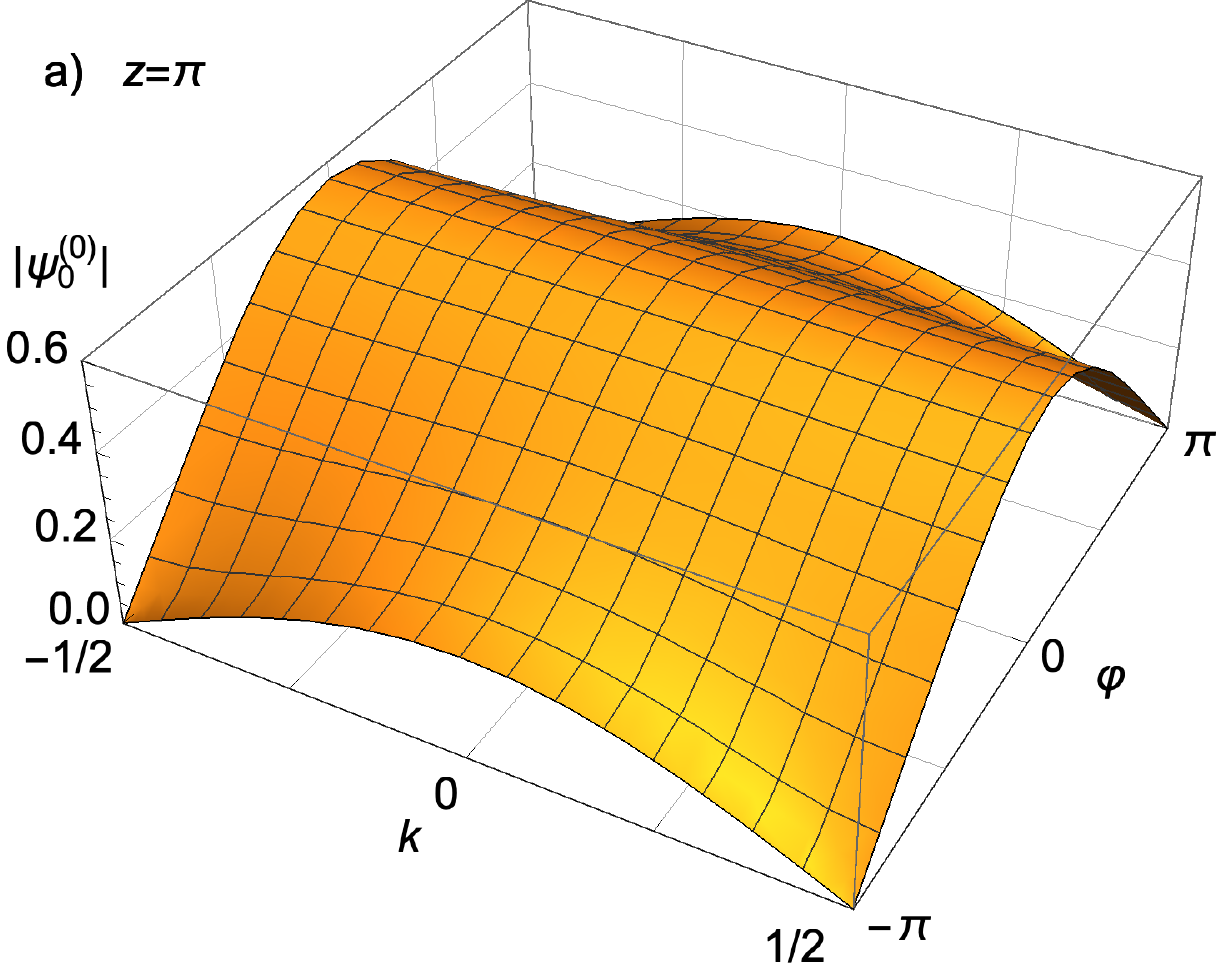} 
\includegraphics[height=4.5cm]{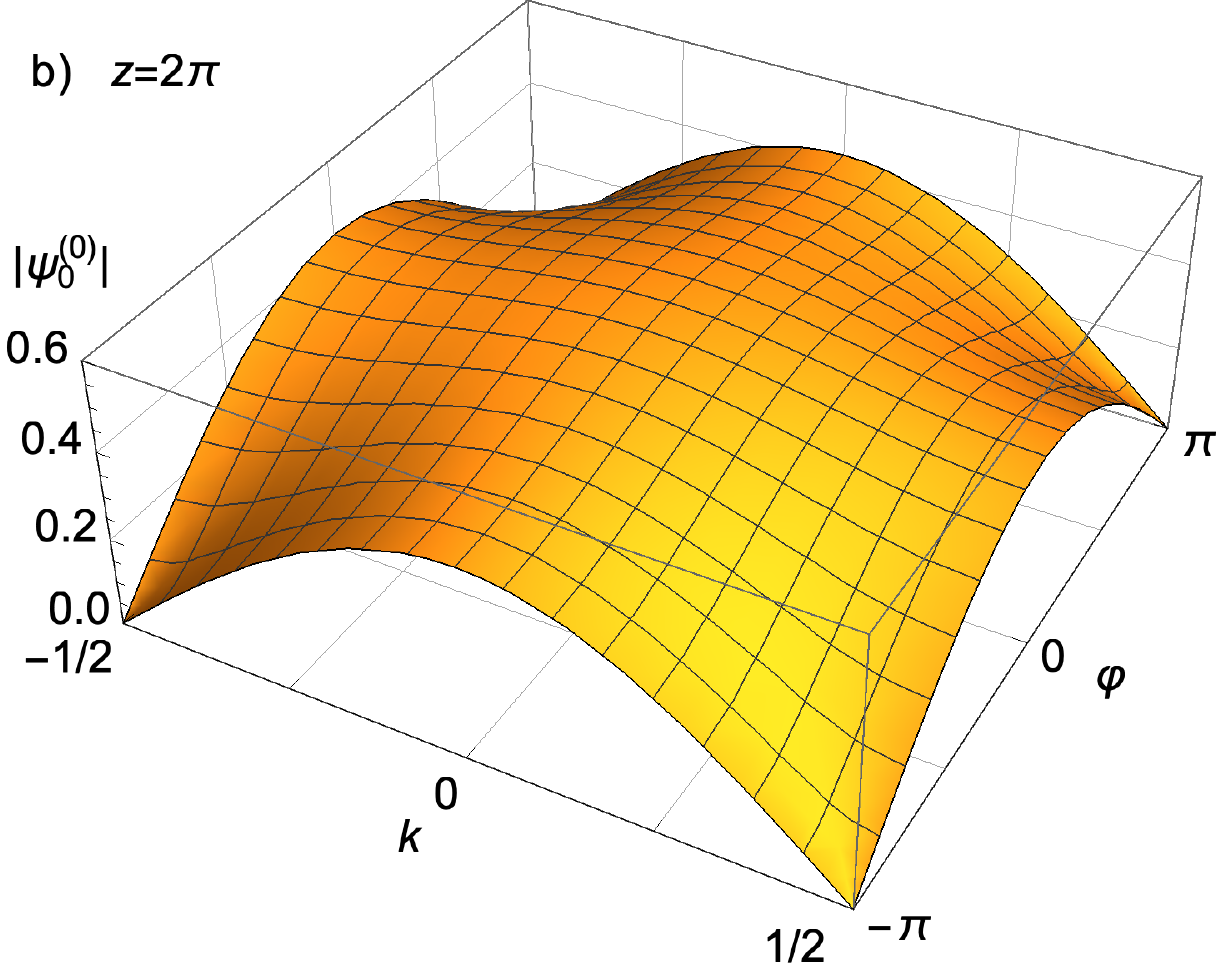} 
\includegraphics[height=4.5cm]{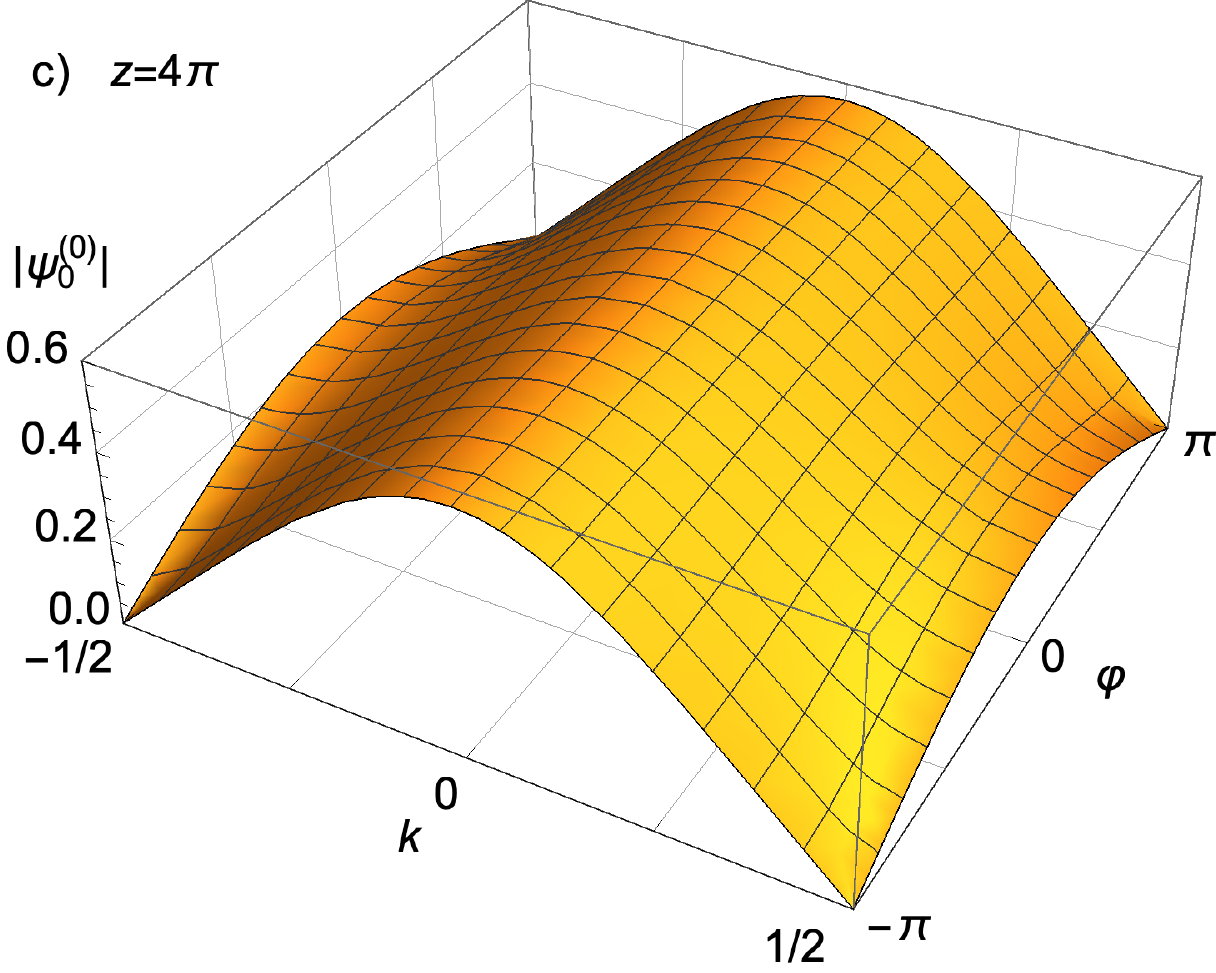} 
\caption{The harmonic oscillator eigenstates in the Zak basis, $|\langle k,\varphi\ket{\psi_0^{(0)}}\!|$, for different values of $z=\sqrt{E_C/E_L}$.   
(a) For  
\mbox{$z<2\pi$}, $\psi_0^{(0)}(k,\varphi)$ is somewhat localised in the modular phase (alternatively, the quasi-flux) coordinate, $\varphi$, but delocalised in $k$.  
(b) For  \mbox{$z=2\pi$}, $\psi_0^{(0)}(k,\varphi)$ is equally delocalised in both coordinates. 
(c) For  \mbox{$z>2\pi$}, $\psi_0^{(0)}(k,\varphi)$ is somewhat localised in the modular number coordinate (alternatively, the quasi-charge \cite{likharev1985theory}), $k$, but delocalised in $\varphi$.
\label{Fig:abc}}
\end{figure}

In the Zak basis, we have the useful operator representations \cite{Zak67,Ganeshan16}
\begin{eqnarray}
\bra{k,\varphi} \hat{n} \ket{\psi} &=&  - i \tfrac{\partial }{\partial \varphi} \left\langle k,\varphi | \psi \right\rangle, \\
\bra{k,\varphi} \hat{\phi} \ket{\psi} &=& \big( - i \tfrac{\partial }{\partial k} + \varphi \big) \left\langle k,\varphi | \psi \right\rangle.
\end{eqnarray}

We express eigenstates  of the reduced Hamiltonian ${\hat H}_1({k,\varphi})$ in the Zak basis for mode 1, $\{\ket{l,\theta}_1\}$, and we find that \mbox{$\psi_{m}(l,\! \theta;k,\varphi)\equiv{}_1\!\langle l, \theta \ket{\psi_m({k,\varphi})}_{\!1}$} are eigenfunctions of the differential operator 
\begin{eqnarray}
    {H}_1({k,\varphi}) \! &=& \! E_C \big( \!-\! i \tfrac{\partial}{\partial \theta} \big)^2 \! + \! E_L \big( \!-\! i \tfrac{\partial }{\partial l} + \theta \big)^2 \nonumber \\
    &&{}-\! E_J \! \cos (\theta \!+\! \varphi ) \!-\! E_Q \! \cos ( 2\pi ( l \!-\! k ) ), \quad \label{Eq:Ham2D} 
\end{eqnarray} 
where $\psi_{m}(l,\! \theta;k,\varphi)$ satisfies generalised periodic boundary conditions
\begin{eqnarray}
 \psi_m \left( - 1/2, \theta; {k,\varphi} \right) &=& \psi_m \left( 1/2, \theta;{k,\varphi}\right), \nonumber\\
 \psi_m \left(l, - \pi ; {k,\varphi}\right) &=& e^{2\pi l i} \psi_m \left(l, \pi;{k,\varphi} \right).\nonumber
\end{eqnarray}
In this representation, it is straightforward to  compute the eigensystem of mode 1 numerically.

It is illustrative  to  evaluate the well-known LC harmonic oscillator ground states in the Zak basis, $\hat H_{\rm HO}$. In the phase basis, $\psi_0^{(0)}(\phi)=(\pi z)^{-\frac{1}{4}}e^{-\frac{\phi^2}{2z}}$, where \mbox{$z=\sqrt{E_C/E_L}$}. In the Zak basis, we find that 
\begin{eqnarray}
\psi_0^{(0)}(k,\varphi) \!&=& \! \sum_{j=-\infty}^\infty e^{-2\pi j k i}\psi_0^{(0)}(\varphi-2\pi j)\nonumber \\
\! &=& \! (\pi z)^{-\frac{1}{4}}{e^{-\frac{\varphi ^2}{2 z}} \vartheta _3\big(\pi  k \! + \! i \pi  \varphi /z,e^{-\frac{2 \pi ^2}{z}}\big)}, \quad \quad
\end{eqnarray}
where $\vartheta_3(u,q)\equiv 1+2\sum_{j=1}^\infty q^{j^2}\cos(2uj)$ is an elliptic theta function. 
The Zak basis highlights the relative localisation of the ground state in $k$ or $\varphi$.  
We plot this function for $z=\pi, 2\pi,$ and $4\pi$ in \cref{Fig:abc}.   When the energy scales for the kinetic and potential terms are balanced, at $z=2\pi$, the ground state is equally delocalised in each coordinate. At this point, the JJ and QPS renormalisation scale factors are equal, i.e.,\ $E_{J}'/E_J=E_Q'/E_Q$.

\section{Evaluating $\hat{\mathcal A}_n$  in the ground state eigenbasis manifold $\{ \ket{\Psi_{0;k,\varphi}}_{1,2} \}$} \label{appen:matrixelement}

Here we compute $ \tensor[_{\!1\!,2}]{\bra{\Psi_{0;k,\varphi}} \!{\hat {\mathcal A}_n}\! \ket{\Psi_{0;k',\varphi'}\!} }{_{\!1\!,2}}$,
\begin{eqnarray}
\tensor[_{1\!,2}]{\bra{\Psi_{0;k,\varphi}} \!{\hat {\mathcal A}_n}\! \ket{\Psi_{0;k',\varphi'}\!} }{_{\!1\!,2}} & = & 2E_C \tensor[_{1\!}]{\bra{\psi_{0}(k,\varphi)} \! \hat n_1 \! \ket{\psi_{0}(k ,\varphi)}  }{_{1}} \nonumber \\
&& \times \delta(k \! - k') \delta(\varphi \! - \! \varphi'). \label{eqn:An_n1}
\end{eqnarray}
We expand $\ket{\psi_{m}(k,\varphi)}_{1}$ to  first order in perturbation theory, and for notational convenience  we define the matrix element
\mbox{$
\llbracket \hat B \rrbracket_{jm}  \!=\! \! \tensor[_{1\!}]{ \bra{\psi_{j}^{(0)}}  \hat B  \ket{\psi_{m}^{(0)}}  } {_{\!1}} 
$} for an operator $\hat B$, so
\begin{equation}
 \ket{\psi_{m}(k,\varphi)}_{\!1} = \ket{\psi_{m}^{(0)}}_{\!1} + \!\!\sum\limits_{j \ne m}\!\! \tfrac{ \llbracket \hat V (k,\varphi) \rrbracket_{jm}}{\Delta E^{(0)}_{mj}} \ket{\psi_{j}^{(0)}}_{\!1}, \label{eqn:NonIdeal_State_Perturbation}
\end{equation}
where \mbox{$\Delta E^{(0)}_{mj}  =  E_m^{(0)} -  E_{j}^{(0)}= \hbar \Omega(m-j)$}.  Then, 
\begin{eqnarray} 
\hspace{-0.5cm} \tensor[_{\!1\!}]{\bra{\psi_{0}(k,\varphi)} \! \hat n_1 \! \ket{\psi_{0}(k,\varphi)}  }{_{\!1\!}}  & = &  \nonumber\\
&&{}\hspace{-3.4cm}  \llbracket\hat n_1\rrbracket_{00}\!+\!    2\!  \! \sum\limits_{m\ne 0} \!\!\mathrm{Re}\big(\tfrac{ \llbracket\hat n_1\rrbracket_{0m} \llbracket \hat V (k,\varphi) \rrbracket_{m0}  }{\Delta E^{(0)}_{0m}}\!\big) 
     \!+\!O(|\tfrac{V}{\Delta E}|^2).{}\label{eqn:n1_matrix}
 \end{eqnarray}

 We write the coordinate $\hat{n}_1$ in terms of creation and annihilation operators $\hat a_1^\dag$ and $\hat a_1$, i.e.,\ \mbox{$\hat{n}_1 =i (\hat a_1^\dag - \hat a_1)/\sqrt{2z} $}. Since $\hat a_1^\dag$ and $\hat a_1$ are off-diagonal in the  harmonic oscillator eigenbasis, we find \mbox{$\llbracket\hat n_1\rrbracket_{0m}=-i\,\delta_{m,1}/{\sqrt{2z}}$}.  Ignoring terms $O(|\tfrac{V}{\Delta E}|^2)$ and higher, we find
\begin{equation}
    \tensor[_{\!1\!}]{\bra{\psi_{0}(k,\varphi)} \! \hat n_1 \! \ket{\psi_{0}(k,\varphi)}  }{_{\!1}} = 2 \,\mathrm{Re} \Big( \tfrac{\llbracket\hat n_1\rrbracket_{01}  \llbracket \hat V (k,\varphi) \rrbracket_{10}}{\Delta E^{(0)}_{01}}    \Big).
\end{equation}

Evaluating the matrix element for $\hat V(k,\varphi)$ gives
\begin{eqnarray}
\llbracket \hat V(k,\varphi) \rrbracket_{10} \! &=&\!
- E_Q \, \tensor[_{1}]{ \bra{\psi_{1}^{(0)}}  \cos \!\big(2\pi (\hat n_1  -  k)\big)   \ket{\psi_{0}^{(0)}}  } {_{1}}  
\nonumber\\
&&-  E_J \,\tensor[_{1}]{ \bra{\psi_{1}^{(0)}}  \cos (\hat \phi_1 +  \varphi) \! \ket{\psi_{0}^{(0)}}  } {_{1}},\nonumber \\
&=&  \!- E_Q \int_{-\infty}^{\infty} \!\!\!dn \cos\!\big(2\pi (n -  k)\big)\, \psi_1^{(0)}\!(n)^{*} \,  \psi_{0}^{(0)}\!(n) 
\nonumber\\
&& -  E_J \!\int_{\infty}^{\infty} \!\!d\phi   \cos (\phi  +  \varphi)\, \psi_1^{(0)}\!(\phi) ^{*} \, \psi_{0}^{(0)}\!(\phi), \nonumber \\
&=&\! {e^{-\frac{z}{4}} \! \sqrt{\tfrac z2} } E_J \sin (\varphi) \!  - \! {i \pi  e^{-\tfrac{\pi^2}{z}}\!\sqrt{\tfrac2z}}E_Q \sin (2\pi k)    . \nonumber
\end{eqnarray}
We  substitute the results into Eq. \eqref{eqn:An_n1}, and noting that $\hbar\Omega z=2E_C$, we obtain Eq. \eqref{eq:para_An}.

 The computation for $\tensor[_{\!1\!,2}]{\bra{\Psi_{0;k,\varphi}} \!{\hat {\mathcal A}_{\phi}}\! \ket{\Psi_{0;k',\varphi'}\!} }{_{\!1\!,2}}$ is implemented in a similar manner.

\section{Resonator based charge syndrome measurement and qubit readout} \label{sec:resonator}

We recall that the external charge $n_x$ typically drifts over time, which is a source of noise for the dualmon. It is thus necessary to detect and correct such type of charge noise. A possible way to do so is as follows. Firstly, from Eqs. \eqref{eq:H_sys_smallnoise}, \eqref{eq:projectednoiseAn}, and \eqref{eq:projectednoiseAphi} it follows that one can couple to the system operators
\begin{subequations}
\begin{eqnarray}
    \hat{\mathcal {A}}_{n}' &=& 2\pi E_Q' \sin (2\pi \hat{n}_2), \\
    \hat{\mathcal {A}}_{\phi}' &=& E_J' \sin(\hat{\phi}_2),
\end{eqnarray}
\end{subequations}
by capacitively or inductively coupling to the dualmon circuit, respectively.  Focusing on an encoding into the ground state $\ket{0,0}$ and one saddle point $\ket{0,\pi}$, the operator $\hat {\mathcal A}_n'$ can act as syndrome for charge noise. This follows from the fact that a shift in external charge $n_x \to n_x - \varepsilon$ can equivalently be interpreted as a shift in $k$. Since $\sin(2\pi \hat n_2)\ket{\varepsilon,\varphi} = \sin(2\pi \varepsilon)\ket{\varepsilon,\varphi}$, a shift with magnitude $|\varepsilon| < 1/4$ can in principle be detected by non-destructively measuring $\hat {\mathcal A}_n'$.

Also, a measurement of $\hat {\mathcal A}_\phi'$ interestingly can be used to readout the logical qubit state. Particularly, we can set the external flux to $\phi_x=-\pi/2$ such that \mbox{$\hat {\mathcal A}_\phi' \to \hat {\mathcal A}_{\phi + \pi/2}' =E_J'\cos(\hat\phi_2)$}. As $\cos(\hat \phi_2) \ket{0,0} = \ket{0,0}$ and $\cos(\hat \phi_2)\ket{0,\pi} = -\ket{0,\pi}$, a measurement of this operator corresponds to a logical basis measurement. Moreover, the $\hat{\mathcal A}'_{\phi}$-measurement is robust to small shifts in $\varphi$.

A potential approach to non-destructively measuring the operators $\hat {\mathcal A}_n'$ and $\hat {\mathcal A}_{\phi}'$ is to couple the dualmon circuit to a resonator. The resulting coupling is of the similar form to ~\cref{eq:H_sys_smallnoise} with the replacements \mbox{$n_x(t) \to n_x(t) + \hat n_r$,  $\phi_x(t) \to \phi_x(t) + \hat \phi_r$}, for capacitive or inductive coupling, respectively. Here $\hat n_r \sim \hat a^\dag_r + \hat a_r$ is the charge bias, and $\hat \phi_r \sim i\hat a^\dag_r - i\hat a_r$ the flux bias due to the resonator, with $\hat a_r$ the resonator annihilation operator. The non-demolition measurement of interest then may be performed by modulating the coupling strength, based on the longitudinal readout scheme proposed in Ref. \cite{Didier2015}. A similar scheme might even be employed to enact resonator induced one-qubit and two-qubit phase gates \cite{Royer2017}. A more detailed study of resonator based readout and control will be the subject of future work.

\section{Coupling to a waveguide} \label{append:waveguide}

\begin{figure}[t]
    \centering
    \includegraphics[scale=0.65]{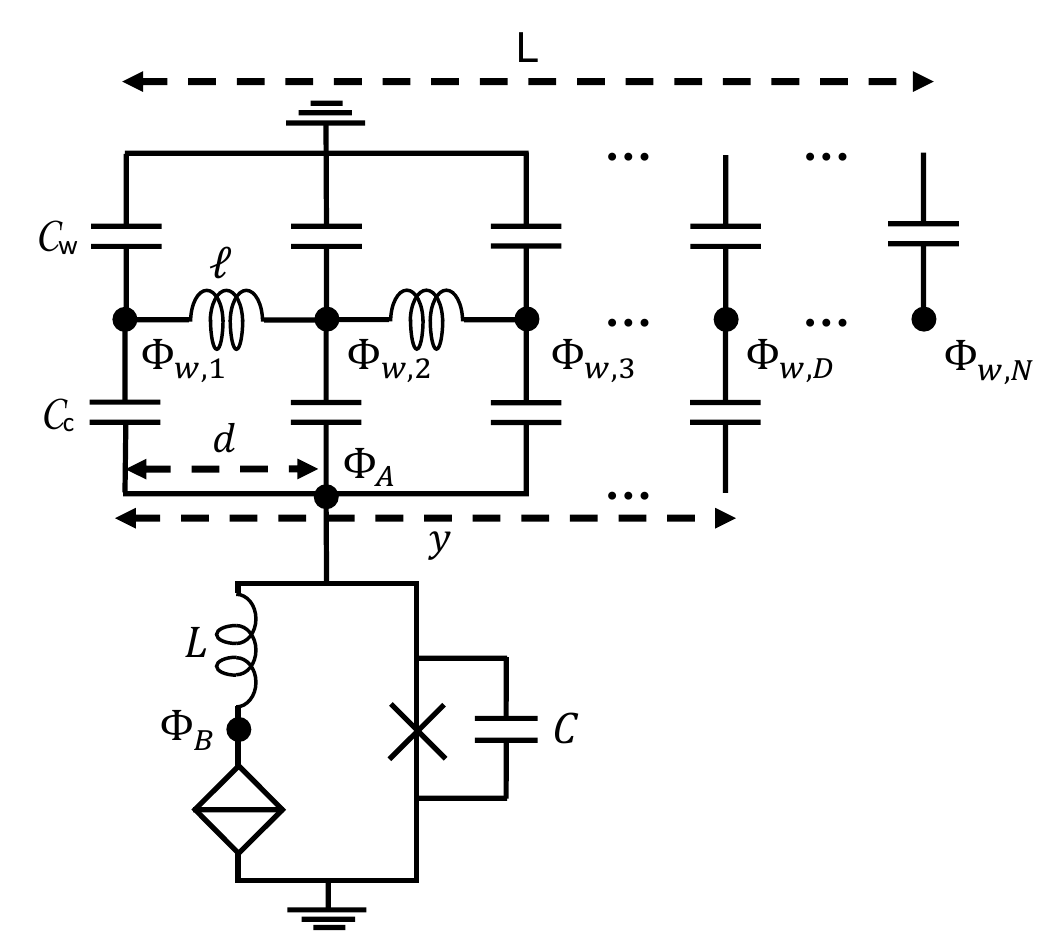}
    \caption{Microscopic model of the capacitive coupling of the realistic dualmon circuit to the waveguide.}
    \label{fig:micro}
\end{figure}

\subsection{Hamiltonian derivation}

Figure \ref{fig:micro} shows the microscopic model of the capacitive coupling between the realistic dualmon circuit and the waveguide. The waveguide of length $\textsf{L} $ is discretely decomposed into  unit cells of length $d$. The number of unit cells is $N=\textsf{L}/d$, each of which has an inductance $\ell$ and a ground capacitance $\mathcal{C}_w$; the waveguide inductance and capacitance per unit length are $\bar \ell =\ell /d$ and $\bar{\mathcal{C}}_w = \mathcal{C}_w/d$. The coupling capacitance is $C_{c}$. This capacitance has a length of $y$ across $D = y/d$ unit cells of the waveguide, yielding a distributed coupling capacitance $\mathcal{C}_c = C_{c}/D$ per unit cell. We also assume that the size of the coupling capacitance is very small compared to that of the waveguide, i.e., $y \ll  \textsf{L} $.

Starting from  the classical equations of motion, the Lagrangian of the net circuit is given by
\begin{eqnarray}
\mathcal{L}_{\mathrm {tot}} &=& \dot \Phi_B Q_Q (\dot \Phi_B) + \tfrac{2e V_c}{2\pi} \cos \big( \tfrac{2\pi}{2e} Q_Q (\dot \Phi_B) \big) + \tfrac{C}{2} \dot \Phi_A^2 \nonumber \\
&& +  \sum_{j=1}^{D} \tfrac{\mathcal{C}_c}{2} (\dot \Phi_{w,j} - \dot \Phi_A)^2 +  \sum_{j=1}^{N} \tfrac{\mathcal{C}_w}{2} \dot \Phi_{w,j}^2 \nonumber \\
&& + \tfrac{I_c \Phi_0}{2\pi} \cos \big( \tfrac{2\pi}{\Phi_0} \Phi_A \big) - \tfrac{1}{2L} (\Phi_A - \Phi_B)^2 \nonumber \\
&& - \sum_{j=1}^{N} \tfrac{1}{2\ell} (\Phi_{w,j+1} - \Phi_{w,j})^2,
\end{eqnarray}
where 
\begin{eqnarray}
Q_Q (\dot \Phi_B ) = \tfrac{2e}{2\pi} \arcsin (\dot \Phi_B /V_c).
\end{eqnarray}
The momenta $Q_A, Q_B, Q_{w,1}, \dots, Q_{w,N}$ are respectively
\begin{eqnarray}
Q_A \!&=& \! C \dot \Phi_A \!+\!  \sum_{j=1}^{D} \mathcal{C}_c (\dot \Phi_A \!- \! \dot \Phi_{w,j}), \label{eq:nonideal_QA} \\
Q_B \!&=& \! Q_Q (\dot \Phi_B), \label{eq:nonideal_QB} \\
Q_{w,j} \!&=& \! \mathcal{C}_c (\dot \Phi_{w,j} \! - \! \dot \Phi_A ) \!+\! \mathcal{C}_w \dot \Phi_{w,j}; \hspace{0.2cm} j\!=\!1, \dots, D, \label{eq:nonideal_Qj1D} \\
Q_{w,j} \!&=& \! \mathcal{C}_w \dot \Phi_{w,j}; \hspace{2.617cm}  j\!=\!D\!+\!1, \dots, N. \label{eq:nonideal_QjD+1N} \quad \quad 
\end{eqnarray}
The Hamiltonian is then
\begin{eqnarray}
\mathcal H_{\mathrm {tot}} \! &=& \! - \tfrac{2e V_c}{2\pi} \cos\big( \tfrac{2\pi}{2e} Q_Q (\dot \Phi_B) \big) \! - \! \tfrac{I_c \Phi_0}{2\pi} \cos \big ( \tfrac{2\pi}{\Phi_0} \Phi_A \big) \nonumber \\
&& + \tfrac{1}{2L} (\Phi_A - \Phi_B)^2 \! + \! \sum_{j=1}^{N} \tfrac{1}{2\ell} (\Phi_{w,j+1} - \Phi_{w,j} )^2 \nonumber \\
&& + \tfrac{C}{2} \dot \Phi_A^2 \! + \! \sum_{j=1}^{D} \tfrac{\mathcal{C}_c}{2} (\dot \Phi_{w,j} \! - \! \dot \Phi_A)^2 \! + \!  \sum_{j=1}^{N} \tfrac{\mathcal{C}_w}{2} \dot \Phi_{w,j}^2. \quad \quad \label{eq:nonideal_H}
\end{eqnarray}
 Eqs. (\ref{eq:nonideal_QB}) and (\ref{eq:nonideal_QjD+1N}) show
\begin{eqnarray}
\dot \Phi_B &=& V_c \sin \big( \tfrac{2\pi}{2e} Q_B \big), \label{eq:nonideal_dotPhiB} \\
 \dot \Phi_j &=& Q_{w,j} / \mathcal{C}_w; \hspace{0.2cm} j = D+1, \dots, N. \label{eq:nonideal_dotPhijN+1D}
\end{eqnarray}
Thus, to express $\mathcal H_{tot}$ in terms of coordinates and conjugate momenta it requires to find
\begin{eqnarray}
\dot \Phi_A &=& \dot \Phi_A (Q_A, Q_{w,1}, \dots, Q_{w,D}), \\
\dot \Phi_{w,j} &=& \dot \Phi_{w,j} (Q_A, Q_{w,1}, \dots, Q_{w,D}); \hspace{0.2cm} j \!=\!1, \dots, D, \quad \quad \quad
\end{eqnarray}
which,  from Eqs. (\ref{eq:nonideal_QA}) and (\ref{eq:nonideal_Qj1D}), are determined by solving the system of equations
\begin{equation}
    \left\{ \begin{array}{ccc}
         C \dot \Phi_A + \sum_{j=1}^{D} \mathcal{C}_c (\dot \Phi_A - \dot \Phi_{w,j}) &=& Q_A \\
         \mathcal{C}_c (\dot \Phi_{w,1} - \dot \Phi_A) + \mathcal{C}_w \dot \Phi_{w,1} &=& Q_{w,1} \\
          & \vdots &  \\
         \mathcal{C}_c (\dot \Phi_{w,D} - \dot \Phi_A) + \mathcal{C}_w \dot \Phi_{w,D} &=& Q_{w,D}
    \end{array} \right. . \label{eq:nonideal_sysequations}
\end{equation}
We substitute the solutions of \eqref{eq:nonideal_sysequations} (and Eqs. (\ref{eq:nonideal_dotPhiB}) and (\ref{eq:nonideal_dotPhijN+1D}) as well) into Eq. (\ref{eq:nonideal_H}), keep terms to first order of $\mathcal{C}_c$ only, and get
\begin{eqnarray}
 {\mathcal H}_{\mathrm {tot}} \! &=& \! - \tfrac{2e V_c}{2\pi} \cos\big( \tfrac{2\pi}{2e} Q_B \big) - \tfrac{I_c \Phi_0}{2\pi} \cos \big ( \tfrac{2\pi}{\Phi_0} \Phi_A \big) \nonumber \\
&& + \tfrac{1}{2L} (\Phi_A - \Phi_B)^2  + \tfrac{C - D \mathcal{C}_c}{2C^2} Q_A^2 \nonumber \\
&& + \! \sum\limits_{j=1}^{D} \tfrac{\mathcal{C}_w - \mathcal{C}_c}{2 \mathcal{C}_w^2} Q_{w,j}^2 \! + \! \sum\limits_{j=D+1}^{N} \tfrac{1}{2 \mathcal{C}_w} Q_{w,j}^2  \nonumber \\
&& \!+ \! \sum\limits_{j=1}^{N} \tfrac{1}{2\ell} (\Phi_{w,j+1} \!- \! \Phi_{w,j} )^2 \!+\!  \sum\limits_{j=1}^{D} \tfrac{\mathcal{C}_c}{\mathcal{C}_w C} Q_A Q_{w,j}. \quad \quad \quad \label{eq:nonideal_H1}
\end{eqnarray}
Note that for $\mathcal C_c$ very small compared to other capacitances, we have $(C - D \mathcal{C}_c)/(2C^2) \approx 1/(2C) $ and \mbox{$(\mathcal{C}_w - \mathcal{C}_c)/(2\mathcal{C}_w^2) \approx 1/(2\mathcal{C}_w)$}. We then set
\begin{eqnarray}
 & E_{\mathcal{C}_w}\! =\! {(2e)^2}/{(2 \mathcal{C}_w)}, \hspace{0.05cm} E_{\ell} \! = \! {\Phi_0^2}/{(8\pi^2 \ell)},   \hspace{0.05cm} E_{\mathcal{C}_c} \! = \! {\mathcal{C}_c (2e)^2}/{(\mathcal{C}_w C)}  \nonumber\\
&  n_{w,j} = {Q_{w,j}}/{(2e)} , \hspace{0.15cm}  \phi_{w,j} = 2\pi { \Phi_{w,j}}/{\Phi_0} ,  \nonumber
\end{eqnarray}
use notations defined in Appendix \ref{appen:derivationHamil}, and change the coordinates of the dualmon circuit as in \cref{coords} to simplify Eq. \eqref{eq:nonideal_H1} into
\begin{equation}
    \hat {\mathcal H}_{\mathrm {tot}} = \hat{\mathcal H}_{\mathrm {sys}} + \hat {\mathcal H}_{\mathrm {wg}} + \hat {\mathcal H}_{\mathrm {coup}}, \label{eq:nonideal_wholdeH}
\end{equation}
where
\begin{eqnarray}
 \hat {\mathcal H}_{\mathrm {sys}} \! & = & \!  E_C \hat n_1^2+ E_L \hat \phi_1^2 \nonumber \\
 && - E_Q  \cos (2\pi (\hat n_1 \! - \! \hat n_2)) \! - \! E_J  \cos (\hat \phi_1 \! + \! \hat \phi_2), \quad\quad  \\
 \hat {\mathcal H}_{\mathrm {wg}} \! &=& \! \sum_{j=1}^{N} E_{\mathcal{C}_w} \hat n_{w,j}^2 + E_{\ell} (\hat \phi_{w,j+1} - \hat \phi_{w,j})^2, \label{eq:waveguide_bareHam} \\
 \hat {\mathcal H}_{\mathrm {coup}} \! &=& \!  \sum_{j=1}^{D} E_{\mathcal{C}_c} \hat n_1 \hat n_{w,j}. \label{eq:waveguide_couplingHam}
\end{eqnarray}

\subsection{Diagonalisation of the waveguide Hamiltonian}

We diagonalise the bare Hamiltonian of the waveguide in Eq. \eqref{eq:waveguide_bareHam}. We first define the travelling modes
\begin{equation}
    \begin{array}{ccl}
 \bar \phi_s &=& \tfrac{1}{\sqrt{N}} \sum\limits_{j=1}^{N} e^{- 2\pi i j k/N} \hat \phi_{w,j} \\
 \bar n_s &=& \tfrac{1}{\sqrt{N}} \sum\limits_{j=1}^{N} e^{ 2\pi i j k/N} \hat n_{w,j}
\end{array} ,
\end{equation}
which have the inverse
\begin{equation}
    \begin{array}{ccl}
       \hat  \phi_{w,j} &=& \tfrac{1}{\sqrt{N}} \sum\limits_{k=0}^{N-1} e^{ 2\pi i j k/N} \bar \phi_s \\
\hat  n_{w,j} &=& \tfrac{1}{\sqrt{N}} \sum\limits_{k=0}^{N-1} e^{- 2\pi i j k/N} \bar n_s
    \end{array} .
\end{equation}
By this, Eq. \eqref{eq:waveguide_bareHam} becomes
\begin{eqnarray}
\hat{ \mathcal H}_{\mathrm {wg}}  &=& 2 \! \sum_{k=0}^{N/2-1} E_{\mathcal{C}_w} \bar n_s \bar n_{N-s} \! + \! 2E_{\ell} \big (1 \!-\! \cos  (\tfrac{2\pi k}{N}) \big) \bar \phi_s \bar \phi_{N-s}. \nonumber \\
\end{eqnarray}
We then introduce symmetric and antisymmetric modes
\begin{equation}
    \begin{array}{ccc}
       \tilde {n}_{s,+} &=& (\bar n_s + \bar n_{N-s})/\sqrt{2} \\
 \tilde {n}_{s,-} &=& i (\bar n_s - \bar n_{N-s})/\sqrt{2} \\
 \tilde{\phi}_{s,+} &=& (\bar \phi_s + \bar \phi_{N-s})/\sqrt{2} \\
 \tilde \phi_{s,-} &=& - i (\bar \phi_s - \bar \phi_{N-s})/\sqrt{2}
    \end{array} ,
\end{equation}
where $0< k< N/2-1$. It follows that
\begin{eqnarray}
\hat{ \mathcal H}_{\mathrm {wg}} &=&\sum_{s=0}^{N/2-1} E_{\mathcal C_w} (\tilde n_{s,+}^2 + \tilde n_{s,-}^2) \nonumber \\
 && + 2E_{\ell} \big (1- \cos \tfrac{2\pi s}{N} \big) (\tilde \phi_{s,+}^2 + \tilde \phi_{s,-}^2) .
\end{eqnarray}
We define
\begin{equation}
    \begin{array}{ccc}
         \tilde n_{s,\pm} &=& \big( \frac{2 E_{\ell}(1-\cos 2\pi s/N)}{E_{\mathcal C_w}} \big)^{1/4} \tilde{\tilde n}_{s,\pm} \\
         \tilde \phi_{s,\pm} &=& \big( \frac{E_{\mathcal C_w}}{2 E_{\ell} (1- \cos 2\pi s/N)} \big)^{1/4} \tilde{\tilde \phi}_{s,\pm} \\
        \hat  a_{s,\pm} &=& (-i \tilde{\tilde \phi}_{s,\pm} +  \tilde{\tilde n}_{s,\pm})/\sqrt{2} \\
 \hat   a_{s,\pm}^\dag &=& ( i \tilde{\tilde \phi}_{s,\pm} + \tilde{\tilde n}_{s,\pm})/\sqrt{2} \\
     \hbar \omega_s &=& \sqrt{2 E_{\mathcal C_w} E_{\ell} (1- \cos (2\pi s/N))}
    \end{array},
\end{equation}
and
take the limit $N \to \infty$ to achieve
\begin{eqnarray}
    \omega_s &=& \pi s/ \big(\textsf{L} \sqrt{\bar{\mathcal C}_w \bar \ell}\big), \\
   \hat{ \mathcal H}_{\mathrm{wg}} &=& \sum_{s=0}^{\infty} \hbar \omega_s (\hat a^{\dag}_{s,+} \hat a_{s,+}  + \hat a^{\dag}_{s,-} \hat a_{s,-}  ). \label{eq:nonideal_Hwg}
\end{eqnarray}

\subsection{The coupling Hamiltonian}
We rewrite the coupling Hamiltonian \eqref{eq:waveguide_couplingHam}
in terms of $\hat a_{s,\pm}$ and $\hat a_{s,\pm}^\dag$. Concretely, 
\begin{widetext}
\begin{eqnarray}
 \hat {\mathcal{H}}_{\mathrm {coup}} &=& \sum\nolimits\nolimits_{j=1}^{D} E_{\mathcal C_c} \hat n_1 \hat n_{w,j} \nonumber \\
 &=& \sum\nolimits\nolimits_{j=1}^{D} E_{\mathcal C_c} \hat n_1 \, \tfrac{1}{\sqrt{N}} \sum\nolimits\nolimits_{s=0}^{N-1} e^{- 2\pi i j s/N} \bar n_s \nonumber \\
 &=& \sum\nolimits\nolimits_{j=1}^{D} E_{\mathcal C_c} \hat n_1 \, \tfrac{1}{\sqrt{N}}  \sum\nolimits\nolimits_{s=1}^{N/2-1} \big( e^{- 2\pi i j s/N} \bar n_s +  e^{2\pi i j s/N} \bar n_{N-s} \big) \nonumber \\
 &=& \sum\nolimits\nolimits_{j=1}^{D} E_{\mathcal C_c} \hat n_1 \, \tfrac{1}{\sqrt{N}}  \sum\nolimits\nolimits_{s=1}^{N/2-1} \big( e^{- 2\pi i j s/N} (\tilde n_{s,+} - i \tilde n_{s,-})/\sqrt{2}  +  e^{2\pi i j s/N} (\tilde n_{s,+} + i \tilde n_{s,-})/\sqrt{2} \big) \nonumber \\
 &=& \sum\nolimits\nolimits_{j=1}^{D} E_{\mathcal C_c} \hat n_1 \, \sqrt{\tfrac{2}{N}}  \sum\nolimits\nolimits_{s=1}^{N/2-1} \big ( \cos (2\pi j s/N) \tilde n_{s,+} - \sin (2\pi j s/N) \tilde n_{s,-}  \big) \nonumber \\
 &=& E_{\mathcal C_c} \hat n_1 \, \sqrt{\tfrac{2}{N}}  \sum\nolimits\nolimits_{s=1}^{N/2-1} \int_{0}^{y} \tfrac{dx}{d} \big ( \cos (2\pi sx/\textsf{L}) \bar n_{s,+} - \sin (2\pi s x/ \textsf{L}) \tilde n_{s,-}  \big) \hspace{0.5cm} (x=jd) \nonumber \\
 &=& E_{\mathcal C_c} \hat n_1 \, \sqrt{\tfrac{2}{N}}  \sum\nolimits\nolimits_{s=1}^{N/2-1} \left( \tfrac{\textsf{L}}{2\pi s  d} \sin(2\pi s y/ \textsf{L}) \tilde n_{s,+} - \tfrac{\textsf{L}}{\pi s d} \sin^2 (\pi s y/\textsf{L}) \tilde n_{s,-} \right) \nonumber \\
&=& E_{\mathcal C_c} \hat n_1 \, \sqrt{\tfrac{2}{N}}  \sum\nolimits\nolimits_{s=1}^{N/2-1} \left( \tfrac{y}{d} \tilde n_{s,+} - \tfrac{\pi s y^2}{d\textsf{L}} \tilde n_{s,-} \right) \hspace{0.3cm} (\mathrm{the\: second\; term\; will\; be\; dropped\; in\; the\; limit\; } \textsf{L} \to \infty) \nonumber \\
&=& \frac{C_c (2e)^2}{\mathcal C_w C} \hat n_1 \, \sqrt{\tfrac{2}{N}}  \sum\nolimits_{s=1}^{N/2-1} \tilde n_{s,+} \hspace{0.3cm} (\mathrm{note\; that\;}  E_{\mathcal{C}_c} = {\mathcal{C}_c (2e)^2}/{(\mathcal{C}_w C)} \mathrm{\, and\,} \mathcal C_c {y}/{d} = \mathcal{C}_c D = C_c)  \nonumber \\
&=& \frac{C_c (2e)^2}{\mathcal C_w C} \hat n_1 \, \sqrt{\tfrac{2}{N}}  \sum\nolimits_{s=1}^{N/2-1} \left( \frac{2 E_{\ell}(1-\cos (2\pi s/N))}{E_{\mathrm C_w}} \right)^{1/4} \tilde{\tilde n}_{s,+} \nonumber \\
&=& \frac{C_c (2e)^2}{\mathcal C_w C} \hat n_1 \, \sqrt{\tfrac{2}{N}}  \sum\nolimits_{s=1}^{N/2-1} \left ( \frac{4 E_{\ell} (\pi s d/\textsf{L})^2}{E_{\mathcal C_w}} \right)^{1/4} \tilde{\tilde n}_{s,+} \nonumber \\
&=& \frac{C_c (2e)^2}{\bar{\mathcal C}_w C} \hat n_1 \, \sqrt{\tfrac{2}{\textsf{L}}}  \sum\nolimits_{s=1}^{N/2-1} \left( \frac{E_{\ell}}{E_{\mathcal C_w}} \right)^{1/4} \left( \frac{2\pi s }{\textsf{L}} \right)^{1/2} \tilde{\tilde n}_{s,+} \nonumber \\
&=& \sum\nolimits_{s=0}^{\infty} g_{s} ( \hat a_{s,+}^\dag + \hat a_{s,+} )\hat n_1, \label{eq:nonideal_Hc}
\end{eqnarray}
\end{widetext}
where
\begin{equation}
  g_{s}  = (2e C_c / C) \sqrt{2 \hbar \omega_s / (\bar{\mathcal{C}}_w \textsf{L})}.
\end{equation}
Since $\hat{\mathcal H}_{\mathrm {coup}}$ couples only modes $(s,+)$ of the waveguide to the dualmon system, in the expression of $\hat{\mathcal H}_{\mathrm {wg}}$ in Eq. \eqref{eq:nonideal_Hwg} we can ignore modes $(s,-)$. For brevity we further shorten the subscript $(s,+)$ into $s$, yielding
\begin{eqnarray}
     \hat{\mathcal H}_{\mathrm {wg}} &=& \sum_s \hbar \omega_s \hat a^\dag_s \hat a_s, \label{eq:appendHwg} \\
    \hat{\mathcal H}_{\mathrm {coup}} &=& \sum_s g_{\omega_s} (\hat a^\dag_s + \hat a_s) \hat n_1. \label{eq:appendHcoup}
\end{eqnarray}
Taking the continuum limit \cite{Fan10}, the two above Hamiltonians  become
\begin{eqnarray}
\hat{\mathcal H}_{\mathrm {wg}} &=& \int_{-\infty}^{\infty} d\omega  \hbar \omega \hat a^\dag (\omega) \hat a(\omega), \\
\hat {\mathcal H}_{\mathrm {coup} } &=& \int_{-\infty}^{\infty} d\omega g(\omega) (\hat a^\dag (\omega) + \hat a (\omega)) \hat n_1, \label{eq:appendHcoupcon}
\end{eqnarray}
 where the lower limit of the frequency $\omega$ has been extended to $- \infty$ \cite{gardiner1985input}, and 
 \begin{equation}
 g(\omega) =  (2e C_c / C) \sqrt{2 \hbar \omega Z_{\mathrm{wg}}/ \pi },
 \end{equation}
 with $Z_{\mathrm{wg}} = \sqrt{\bar \ell/ \bar{\mathcal C}_w}$  the waveguide impedance. The spectral density is given by
\begin{equation}
    J(\omega) = \tfrac{1}{\hbar^2}  \int_{-\infty}^{\infty} d \omega g^2 (\omega) \delta(\omega - \omega_s) =\tfrac{8e^2}{\pi \hbar } \tfrac{C_c^2}{C^2} Z_{\mathrm{wg}} \omega.
\end{equation}
 We can write $J(\omega)=\nu \hbar \omega$ with
\begin{equation}
\nu = \tfrac{8 e^2}{\pi \hbar^2} \tfrac{C_c^2}{C^2} Z_{\mathrm{wg}}.
\end{equation}

\section{Input-output calculations} \label{append:inout}
 We employ the rotating wave approximation and the Markov approximation \cite{gardiner1985input,Combes17} to simplify the coupling Hamiltonian in Eq. \eqref{eq:appendHcoupcon} to
 \begin{equation}
 \hat {\mathcal H}_{\mathrm {coup} } = \int_{-\infty}^{\infty} d\omega \hbar \sqrt{\gamma/2\pi} (\hat a^\dag (\omega) \hat n_1^{-} + \hat a (\omega) \hat n_1^{+}) ,
 \end{equation}
 where $\hbar \sqrt{ \gamma/2\pi} \equiv g (\omega_D)$ is the coupling strength evaluated at the drive frequency $\omega_D$, and $\hat n^{+}_{1} \, (\hat n^{-}_{1}) $ is the lower (upper) triangularised version of  $\hat n_1$. Note that the Markov approximation is valid in the regime of narrowband interaction, and the operator $\hat n_1^{+} $ in terms of the eigenbasis $\{ \ket{\Psi_{m;k,\varphi}}_{1,2} \}$ is of the form
 \begin{eqnarray}
 \hat n_1^{+} &=& \sum_{m>n} \int_{-1/2}^{1/2} \int_{-\pi}^{\pi} dk d\varphi \tensor[_{1,2}]{\langle \Psi_{m;k,\varphi}| \hat n_1 | \Psi_{n;k,\varphi} \rangle}{_{1,2}} \nonumber \\
 && \times \ket{\Psi_{m;k,\varphi}}_{1,2}\! \bra{\Psi_{n;k,\varphi}}.
 \end{eqnarray} 
  The master equation for the system density operator and the input-output relation are then, using the results in Refs. \cite{gardiner1985input,Combes17}, given by Eqs. \eqref{eq:nonideal_me} - \eqref{eq:inout} in the main text.

\begin{figure}[t]
\includegraphics[height=4.27cm]{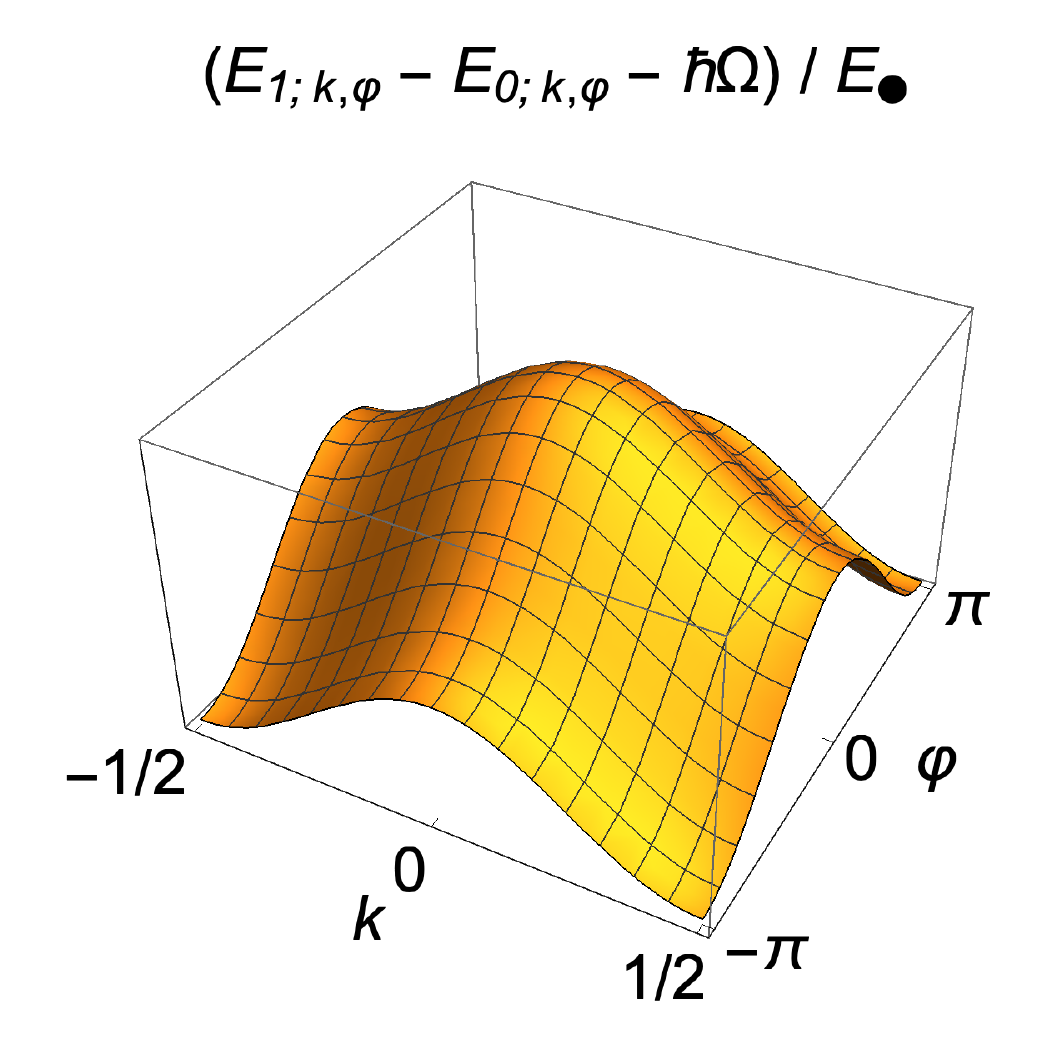} 
\includegraphics[height=4.27cm]{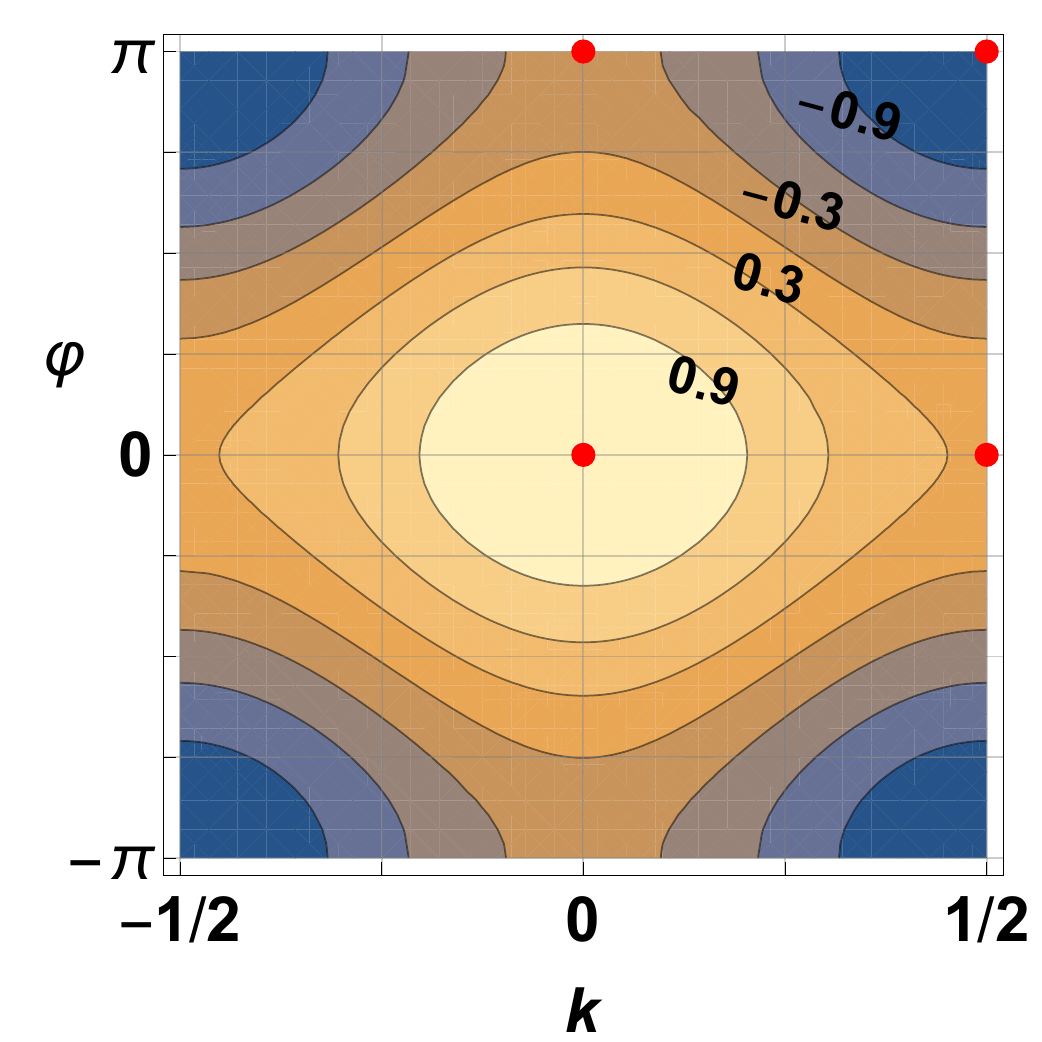} 
\caption{
(Left) and (Right)  Three-dimensional and contour plots of the transition frequency between the the ground state and first excited state manifolds relative to the unperturbed harmonic oscillator frequency, $\hbar \Omega_{k,\varphi}^{(1,0)} - \hbar \Omega \equiv E_{1;k,\varphi}- E_{0;k,\varphi}-\hbar \Omega.$ The transition at the ground state $(k=0,\varphi=0)$ and the maximally excited state $(k=1/2,\varphi = \pi)$ is unique, i.e., maximal and minimal respectively, whereas the transition at each saddle point is degenerate.
For illustrative purposes, plots are drawn with $E_Q=E_J\equiv E_\bullet$, $E_C/E_J=200$ and \mbox{$E_L/E_J=10$}.} \label{Fig:transition}
\end{figure}

The spectrum displayed in \cref{fig:transmission} is obtained by assuming the dualmon state initially in the ground state manifold with a specific pair of $(k,\varphi)$, say, the state $\ket{\Psi_{0;k,\varphi}}_{1,2}$. We then approximate the dualmon as a two-level system with eigensystem
\begin{equation}
    \begin{array}{ll}
         \ket{\Psi_{0;k,\varphi}}_{1,2}\!; & E_{0;k,\varphi} \!=\! \hbar \Omega/2 - E_Q' \cos (2\pi k) - E_J' \cos (\varphi) \\
     \ket{\Psi_{1;k,\varphi}}_{1,2}\!; & E_{1;k,\varphi} \!=\! 3\hbar \Omega/2- E_Q'' \cos (2\pi k) - E''_J \cos (\varphi) 
    \end{array}\!, \nonumber
\end{equation}
where 
\begin{equation}
    \begin{array}{lcl}
         E'_Q &=& e^{-\pi^2/z} E_Q \\
         E'_J &=& e^{-z/4} E_J \\
         E''_Q &=& (1-2\pi^2/z) E'_Q \\
         E''_J &=& (1-z/2) E'_J
    \end{array}.
\end{equation}
The transition frequency, $\hbar \Omega_{k,\varphi}^{(1,0)} = E_{1;k,\varphi} - E_{0;k,\varphi}$, is
\begin{equation}
   \hbar  \Omega_{k,\varphi}^{(1,0)} = \hbar \Omega + \tfrac{2 \pi^2}{z} E'_Q \cos (2\pi k) + \tfrac{z}{2} E'_J \cos (\varphi). \label{eq:transition}
\end{equation}
In \cref{Fig:transition}, we plot the variation of $\Omega_{k,\varphi}^{(1,0)}$ with respect to $(k,\varphi)$. The transition is unique at the ground state $(k=0,\varphi=0)$ and the maximally excited state $(k=1/2,\varphi=\pi)$ only, being maximal and minimal respectively. 

We move to the frame rotating at the drive frequency $\omega_D$, making the master equation \eqref{eq:nonideal_me} time-independent. We then solve for the steady-state solution of this master equation, and compute the transmission. Relevant parameters for plotting are chosen as $E_Q =E_J,\hbar \Omega = 40\sqrt{5}\, E_J, z=\sqrt{20},$ $  \gamma =0.01\, \omega_D, $ and $\hbar \alpha^2 =0.1 \,E_J$.
Noteworthily, a time-dependent treatment with higher truncation for the transmission problem yields a considerably similar result.

\section{Scanning over bias charge and flux} \label{append:scan}

We recall that the spectrum shown in \cref{Fig:NonIdeal_BandStructure} is computed by numerically solving the eigensystem of mode 1 described by the Hamiltonian $\hat H_1 (k,\varphi)$ in \cref{Eq:NonIdeal_HamiltonianMode1} in the absence of flux and charge biases. When the two biases are nonzero, the Hamiltonian of mode 1 is
\begin{eqnarray}
\hat H_1 (k,\varphi;n_x,\phi_x) \!&=& \!E_C (\hat n_1 \! + \! n_x)^2 \!+ \! E_L (\hat \phi_1 \! - \! \phi_x)^2 \nonumber \\
&& - E_Q \cos (2\pi (\hat n_1 \!-\! k))  \! - \! E_J \cos (\phi_1 \! + \! \varphi). \nonumber \\
\end{eqnarray}
It can be checked that under the unitary transformation 
\begin{equation}
\hat U(n_x,\phi_x)=\exp{(i \hat n_1 \phi_x)} \exp{(i \hat \phi_1 n_x)},
\end{equation}
$\hat H_1 (k,\varphi;n_x,\phi_x)$ becomes
\begin{equation}
\hat U(n_x,\phi_x) \hat H_1 (k,\varphi;n_x,\phi_x) \hat U^\dag (n_x,\phi_x)\! = \!  \hat H_1 (k\!+\!n_x, \varphi \!+\! \phi_x ).
\end{equation}
This implies the spectrum of $\hat H_1 (k,\varphi;n_x,\phi_x)  $ is obtained by shifting the origin of the $\hat H_1 (k,\varphi)$-spectrum to $(-n_x,-\phi_x)$. After turning on the bias parameters the system state (initially assumed to be an eigenstate) is changed into a new eigenstate with a new eigenenergy and new transition frequencies. Hence, scanning over the biases $n_x \in [-1/2,1/2)$ and $\phi_x \in [-\pi,\pi)$ maps the full spectrum of the dualmon system.

\section{Thermal induced dephasing rate}\label{append:thermaldephasing}

We calculate the dephasing rates due to thermal excitations and relaxations for two different cases, that is, a qubit that is encoded in an ordinary two-level system, and a qubit that is encoded in the ground state manifold analogous to the proposed dualmon qubit. 

The master equation for the standard two-level-system case is
\begin{equation}
\dot   \rho = - \tfrac{i}{\hbar} [\hat H, \rho] + \gamma_{-} \mathcal{D}[\hat \sigma_-] \rho +\gamma_{+} \mathcal D [\hat \sigma_+] \rho,
\end{equation}
where
\begin{equation}
    \begin{array}{ccl}
        \hat H & = &  \tfrac{1}{2}\hbar  \omega (   \ket{e} \bra{e} -\ket{g} \bra{g}) \\ 
       \hat \sigma_- &=& \ket{g} \bra{e} \\
       \hat  \sigma_+ &=& \ket{e} \bra{g}
        \end{array},
\end{equation}
 and $\gamma_- (\gamma_+)$ is the relaxation (excitation) rate. From the master equation, we find  the evolution equation for the off-diagonal element $\rho_{eg}$ as follows
\begin{equation}
    \dot \rho_{eg} = - i \omega \rho_{eg} - \tfrac{1}{2} (\gamma_+ + \gamma_-) \rho_{eg},
\end{equation}
showing the dephasing rate is proportional to the sum of both the relaxation and excitation rates.

With regards to the qubit encoded in the ground state manifold as in the dualmon device, we assume the system possesses a ground state manifold consisting of $\ket{g_1}$ and $\ket{g_2}$ and an excited state manifold with $\ket{e_1}$ and $\ket{e_2}$. By analogy to the coupling operator $\hat n_1$ in \cref{subsec:thermal}, we further assume that the coupling of the system to a heat bath does not induce transitions within each manifold, but allows selected transitions between different manifolds, namely, $\ket{g_j} \leftrightarrow \ket{e_j} $ (not including $\ket{g_j} \leftrightarrow \ket{e_{j'}}$ for $j \ne j'$). In this case, the relevant master equation is
\begin{equation}
     \dot \varrho = - \tfrac{i}{\hbar } [\hat {\mathcal H}, \varrho] + \sum_{j=1}^{2} \Big( \gamma_{-}^{(j)}  \mathcal{D} \big [\hat{\sigma}_{-}^{(j)}\big] \varrho +  \gamma_{+}^{(j)}  \mathcal{D} \big [\hat{\sigma}_{+}^{(j)} \big] \varrho \Big),
\end{equation}
where 
\begin{equation}
    \begin{array}{ccl}
         \hat{\mathcal H} &=& \sum_{j=1}^{2}  \tfrac{1}{2} \hbar \omega_j (   \ket{e_j}  \bra{e_j} -\ket{g_j}  \bra{g_j} ) \\
          \hat { \sigma}_-^{(j)} &=& \ket{g_j} \bra{e_j} \\
       \hat{ \sigma}_+^{(j)} &=& \ket{e_j} \bra{g_j}
    \end{array} ,
\end{equation}
and $\gamma_-^{(j)} $ and $\gamma_{+}^{(j)}$ are respectively the relaxation and excitation rates between $\ket{g_j}$ and $\ket{e_j}$. The time evolution for the off-diagonal element $\varrho_{g_1 g_2}$, which is related to dephasing of the dualmon qubit, is then
\begin{equation}
   \dot \varrho_{g_1 g_2} =  \tfrac{1}{2} i (\omega_1 - \omega_2) \varrho_{g_1 g_2} - \tfrac{1}{2}( \gamma_+^{(1)} + \gamma_{+}^{(2)} ) \varrho_{g_1 g_2}. 
\end{equation}
This result demonstrates that the dephasing rate for the dualmon qubit is determined by the excitation rates out of the ground state manifold only, remarkably differing from the standard two-level-system case.

\bibliography{qpsjj}

\end{document}